\documentclass[11pt,reqno]{amsart}

\usepackage{amsmath, amsfonts, amssymb, amscd, enumerate, color, relsize}
\usepackage{hyperref}

\theoremstyle{plain}
\newtheorem{theo}{Theorem}[section]

\newtheorem{prop}[theo]{Proposition}

\theoremstyle{definition}

\newtheorem{definition}[theo]{Definition}
\newenvironment{pf}{\noindent{\it Proof. }}{$\square$\par\medskip}

\theoremstyle{plain}

\newtheorem{theorem}[theo]{Theorem}

\theoremstyle{definition}

\renewcommand{\=}{:=}
\newcommand{\beq}{\begin{equation}}
\newcommand{\eeq}{\end{equation}}
\renewcommand{\a}{\alpha}
\renewcommand{\b}{\beta}

\newcommand{\e}{\epsilon}
\newcommand{\ve}{\varepsilon}

\newcommand{\h}{\eta}

\renewcommand{\l}{\lambda}
\renewcommand{\o}{\omega}

\renewcommand{\r}{\rho}
\newcommand{\s}{\sigma}

\newcommand{\x}{\xi}

\newcommand{\D}{\Delta}

\renewcommand{\L}{\Lambda}
\renewcommand{\O}{\Omega}

\newcommand{\bA}{\mathbb{A}}

\newcommand{\bF}{\mathbb{F}}

\newcommand{\bR}{\mathbb{R}}

\newcommand{\gz}{\mathfrak{z}}

\renewcommand\sp{\mathfrak{sp}}

\newcommand{\cC}{\mathcal{C}}
\newcommand{\cD}{\mathcal{D}}
\newcommand{\cE}{\mathcal{E}}

\newcommand{\cI}{\mathcal{I}}

\newcommand{\cL}{\mathcal{L}}

\newcommand{\cS}{\mathcal{S}}

\newcommand{\cU}{\mathcal{U}}
\newcommand{\cV}{\mathcal{V}}
\newcommand{\cW}{\mathcal{W}}

\newcommand{\p}{\partial}

\newcommand{\lra}{\leftrightarrow}

\renewcommand{\square}{\kern1pt\vbox
{\hrule height 0.6pt\hbox{\vrule width 0.6pt\hskip 3pt
\vbox{\vskip 6pt}\hskip 3pt\vrule width 0.6pt}\hrule height0.6pt}\kern1pt}

\DeclareMathOperator{\Span}{Span}

\newcommand{\wt}{\widetilde}
\newcommand{\wh}{\widehat}

\newcommand{\be}{\begin{equation}}
\newcommand{\ee}{\end{equation}}

\def\<#1,#2>{\langle\,#1,\,#2\,\rangle}
\newcommand{\arr}{\begin{array}{rlll}}
\newcommand{\ea}{\end{array}}
\newcommand{\bea}{\begin{eqnarray}}
\newcommand{\eea}{\end{eqnarray}}
\newcommand{\bean}{\begin{eqnarray*}}
\newcommand{\eean}{\end{eqnarray*}}
\newcommand{\vv}{\operatorname{v}}

\def\sideremark#1{\ifvmode\leavevmode\fi\vadjust{%            The remark
\vbox to0pt{\hbox to 0pt{\hskip\hsize\hskip1em%               will appear only
\vbox{\hsize3cm\tiny\raggedright\pretolerance10000%          on the side
\noindent #1\hfill}\hss}\vbox to8pt{\vfil}\vss}}}%           in 3cm
\newcounter{ssig}
\newcounter{ttig}
\newcommand{	\Div}{\operatorname{Div}}

\renewcommand{\vv}{\text{\bf v}}
\newcommand{\vvb}{\text{\bf v}_{\!\!\mathsmaller B}}
\newcommand{\Xb}{X_{\!\!\mathsmaller B}}

\renewcommand{\lra}{\longrightarrow}
\title[Lie algebras of conservation laws]
{Lie algebras of conservation laws of  variational  partial differential equations}
\author[E. Fiorani, S. Germani and A. Spiro]
{Emanuele Fiorani, Sandra Germani and Andrea Spiro}


\thanks{{\it Acknowledgments}. This research was partially supported by the Project MIUR ``Real and Complex Manifolds: Geometry, Topology and  Harmonic Analysis'' and by  GNSAGA of INdAM}

  \address
{\newline Emanuele Fiorani and Andrea Spiro, 
Scuola di Scienze e Tecnologie, Universit\`a di Camerino, Via Madonna delle Carceri 9,
I-62032 Camerino (Macerata),
ITALY\newline
 Sandra Germani,  Via Parini 4,  I-63821 Porto Sant'Elpidio (Fermo), 
ITALY\newline
\phantom{a}}
\email
{emanuele.fiorani@unicam.it, sandragermani27@gmail.com, an\-drea.spi\-ro@unicam.it}\par
\medskip

\begin{document}

\begin{abstract} We establish a   version of the first Noether Theorem, according to which the  (equivalence classes of)  conserved quantities of given
Euler-Lagrange equations  in several independent variables are in  one-to-one correspondence
with the (equivalence classes of) vector fields satisfying   an appropriate pair of geometric conditions, namely: (a)  they  preserve   the class of vector fields tangent to holonomic submanifolds of a jet space; (b)  they leave invariant  the  action,   from which the Euler-Lagrange equations are derived,  modulo   terms  identically vanishing along holonomic submanifolds.
Such correspondence  between symmetries and  conservation laws  is built  on  an  explicit   linear map $\Phi_\a$ from the vector fields satisfying (a) and (b)  into the conserved differential operators,  and  not  into  their divergences  as it occurs  in other   proofs of Noether Theorem.  This map  $\Phi_\a$ is not new:  it is  the  map determined by  contracting symmetries   with a form of  Poincar\'e-Cartan type $\a$ and it  is essentially the same  considered for instance in a paper by Kupershmidt. There it was shown 
that $\Phi_\a$ determines a bijection  between symmetries and conservation laws in a  special form. Here we show that,  if appropriate regularity assumptions are satisfied, 
 {\it any}  conservation law is equivalent to one  that belongs to the image of  $\Phi_\a$, proving  that the corresponding induced map
 $\wt \Phi_\a$ between equivalence classes of symmetries and equivalence classes of  conservation laws is actually a bijection. 
All results are given coordinate-free formulations and  rely   just on basic  differential geometric properties  of  finite-dimensional manifolds.
\end{abstract} 

\maketitle
\setcounter{section}{0}
\setcounter{subsection}{1}
\section{Introduction}
\setcounter{equation}{0}
In a  previous paper (\cite{FS}) it  was  established a new version of the celebrated Noether Theorem on the bijection  
between (equivalence classes of) conservation laws and (equivalence classes of) symmetries of Euler-Lagrange equations for the  
 case of functions of one independent variable. 
 \par
  The main purpose  of that paper
 was to give a self-contained proof of Noether  Theorem,  in a coordinate free formulation and relying      
only   on standard  differential geometric properties   of  finite-dimensional manifolds. An outcome   of 
 this  approach was the realization  of the fact that   Noether's correspondence  between  conservation laws and  symmetries 
 can be  actually determined  by  a  linear map that goes directly  from  the Lie algebra of  infinitesimal symmetries   into 
 the vector space  of constants  of motions,   and  not into the space of their differentials  as it occurs in other proofs of Noether Theorem.  
Such linear map is very simple: it is the map $\Phi_\a$  that sends  an infinitesimal symmetry $X$   into the function $f \= \imath_X \a$, where $\a$  is  a fixed $1$-form,  determined by   the Lagrangian  that gives  the Euler-Lagrange equations. This  $\a$  is  a  generalisation  of the  Poincar\'e-Cartan $1$-form  $\a_H \= p_i dq^i - H dt$  of  Hamiltonian Mechanics. \par
 We have to stress the fact that $\Phi_\a$  is not new:  for instance, it  essentially coincides with the  correspondence between symmetries and  a special class of  conservation laws,   established by Kupershmidt  in \cite{Ku}, Thm. II.5.1,   for Lagrangians and   Euler-Lagrange equations of arbitrary order and for field theories   with an  arbitrary number of  independent variables.  We   emphasise that the main 
result  in \cite{FS}  asserts that  {\it any} conserved quantity is, up to the addition of a trivially conserved quantity,  equivalent to one contained in the image of the mapping $\Phi_\a$. 
This means that  the  induced correspondence $\wt \Phi_\a$ between equivalence classes of symmetries and equivalence classes of conserved quantities is actually a bijection. Kupershmidt, in contrast, only shows that the conserved currents in a particular form, that is those obtained by a particular contraction of a vector field with a form of Poincare-Cartan type, is contained in the image of the mapping $\Phi_\a$. 
  \par
 \smallskip
 In this paper we extend the geometric construction  of \cite{FS} to the general case of  conservation laws  and Euler-Lagrange equations for functions of  $m$ independent variables.  
All notions and  arguments  considered  in \cite{FS}  are  directly  extended to such   general setting.  Differences occur only in  few  points and are due  only to  
 the presence of a higher  number of independent variables.   Actually, during the preparation of this paper, 
we realised that in \cite{FS} the first and third author gave  an incorrect claim, which is here   removed. A detailed erratum for \cite{FS} is given in the appendix. \par
\smallskip
As in the previous paper, Noether's correspondence   between symmetries and conservation laws is established 
 by means of a  linear map $\Phi_\a$, which transforms the elements $X$ of  the Lie algebra of infinitesimal symmetries into   the  conserved $(m-1)$-forms  $\h = \imath_X \a$, where  $\a$ is a fixed $m$-form, called {\it of Poincar\'e-Cartan type}.  Here,  with the expression   ``conserved $(m-1)$-form''  we   mean  an $(m-1)$-form with  components that constitute   a vector valued differential operator  with a divergence that vanishes on the  solutions to the Euler-Lagrange equations. As mentioned above, this linear map $\Phi_\a$ is essentially the same considered in \cite{Ku}, Ch. II.5 and, as in \cite{FS},  our main Theorem \ref{inverseNoether} shows that, under appropriate regularity conditions,  {\it any} conserved quantity is, up to the addition of a trivially conserved $(m-1)$-form,  equivalent to one contained in the image of the mapping $\Phi_\a$.  Due to  this,  we get  that  the  induced map $\wt \Phi_\a$  between equivalence classes of symmetries and equivalence classes of conserved quantities is a true bijection,  improving in this way Kupershmidt's result in full generality.\par
 \smallskip
In order to make as much as possible clear  and explicit all aspects  of innovation of our results,    in \S 2 we overview   Olver's version  of Noether Theorem, which,  at the best of our knowledge,  is  the 
most general and complete variant  of this theorem (see \cite{Ol, Ol1,Ko, Ol2}).   We  then outline our results, pointing out   differences  and similarities with Olver's  and  other variants of Noether Theorem, as  for instance those given  in \cite{Ly, KLR, BCD}.\par
\smallskip
 {\it Structure of the paper.}Ê After  section \S \ref{Olver's proof},  where the reader can find an   outline of all  contents of this paper,  in  \S \ref{geompresent}  we introduce the  main ingredients of our approach, namely the  notions of holonomic forms, variational classes and variational principles for actions defined by  variational classes. In \S \ref{NoetherTheoremsSection},  we prove the
 first and second part of  Noether Theorem: in  the first, we show  that, by contraction with a  fixed   $m$-form of Poincar\'e-Cartan type, any $\cI$-symmetry is associated with   a conserved $(m-1)$-form;  in  the second, we prove that, under appropriate regularity conditions,  this correspondence  can be reversed.  In 
 \S  \ref{Examples}, an explicit   example  of an $m$-form of Poincar\'e-Cartan type is  given. In Appendix,  the above mentioned erratum for \cite{FS} is given.\par
 \medskip
 {\it Acknowledgements.} We are grateful to Franco Cardin and Juha Pohjanpelto for very useful discussions on various aspects of this paper.
\par
\bigskip
\section{An outline  of  the results and  comparisons with previous versions of the Noether Theorem}
\label{Olver's proof} 
\setcounter{equation}{0}
\subsection{A short overview of Olver's version of Noether Theorem}\hfill\par
 Consider a system of partial differential equations of order $k$ of class $\cC^\infty$
 \beq \label{1.1} F_\nu\bigg(x^i, y^j, \frac{\p y^j}{\p x^\ell}, \ldots, \frac{\p^k y^j}{\p x^{\ell_1} \dots \p x^{\ell_k}}\bigg) = 0\ ,\qquad \nu = 1, \dots, N\ ,\eeq
 for $n$ unknown functions $y^j(x^i)$ of $m$ independent variables $x^i$, $ 1 \leq i \leq m$. 
 An  $m$-tuple of smooth differential operators of order $r$ 
 $$P = (P^1, \ldots, P^m)\ ,\qquad  P^\ell = P^\ell\bigg(x^i, y^j, \frac{\p y^j}{\p x^\ell}, \ldots, \frac{\p^{r} y^j}{\p x^{\ell_1} \dots \p x^{\ell_r}}\bigg)$$
 is said
{\it to satisfy  a conservation law for \eqref{1.1}}   if the equation  
\beq \label{1.2} \Div\left(P\bigg(x^i, y^j(x^s), \left.\frac{\p y^j}{\p x^\ell}\right|_{(x^s)}, \ldots, \left.\frac{\p^{r} y^j}{\p x^{\ell_1} \dots \p x^{\ell_r}}\right|_{(x^s)}\bigg)\right) = 0\eeq 
is identically satisfied whenever $y^j(x^s)$ is a solution to   \eqref{1.1}.   If  the $m$-tuple $P$ is identically vanishing on all solutions of \eqref{1.1}, the conservation law is called {\it trivial of the first kind}. If \eqref{1.2} holds  for {\it all} smooth maps $y^j(x^s)$ (not just for  the solutions to \eqref{1.1}),  the conservation law is called  {\it trivial of the second kind}. We shortly call {\it trivial conservation law} any    sum of  such two types of  conservation laws.   \par
\smallskip
Given a non-negative integer $s$,   the   {\it prolongation of \eqref{1.1}  to order $k+s$}  is the p.d.e.'s system  that is determined by    the equations in  \eqref{1.1} together with  their derivatives  up to order $s$. It is therefore a system 
 \beq \label{1.1bis} F^{(s)}_{\nu}\bigg(x^i, y^j, \frac{\p y^j}{\p x^\ell}, \ldots, \ldots, \frac{\p^{r} y^j}{\p x^{\ell_1} \dots \p x^{\ell_r}}, \ldots \frac{\p^{k+s} y^j}{\p x^{\ell_1} \dots \p x^{\ell_{k+s}}}\bigg) = 0\ ,\eeq
 where  now $\nu$ runs from $1$ up to  an appropriate integer $N_{k+s}Ê\geq N_k\= N$, which  depends on  the order $s$ of the prolongation.\par
Now,   it is possible to show that if the map   $F^{(r+1-k)} {=} (F^{(r+1-k)}_\nu)_{1 \leq \nu \leq N_{r+1}}$ locally satisfies an appropriate constant rank condition, then  for any $(r+1)$-th order differential operator of the form $ \Div(P)$, which appear  in a conservation law \eqref{1.2},   there  locally exists a set  of  differential operators of order  $(r+1)$   
 $$Q^{\nu} = Q^{\nu}\bigg(x^i, y^j, \frac{\p y^j}{\p x^\ell}, \ldots, \frac{\p^{r} y^j}{\p x^{\ell_1} \dots \p x^{\ell_r}}\bigg)\ ,\qquad 1 \leq \nu \leq N_{r+1}\ ,$$
such that (see \cite{Ol}, formula (4.27))
\beq \label{1.3}  \Div(P)   =  \sum_{\r = 1}^{N_{r+1-k}} Q^{\nu} F^{(r+1-k)}_{\nu} \ .\eeq
The operators $Q^\nu$ are determined by the operator $\Div(P)$  up to addition of a differential  operator that vanishes identically on all solutions to \eqref{1.1}.  \par
\smallskip
 Assume now  that \eqref{1.1} is a system of Euler-Lagrange equations, that is a system of   equations that characterises  the stationary points,   within the class of  local variations with fixed boundary values, of  a functional 
 \beq\label{1.4}  \cI = \int_{\cU} L\bigg(x^i, y^j, \frac{\p y^j}{\p x^\ell}, \ldots, \frac{\p^{k'} y^j}{\p x^{\ell_1} \dots \p x^{\ell_{k'}}}\bigg) dx^1 \wedge \ldots \wedge dx^m\eeq
 for  some smooth      $L$, usually called    {\it Lagrangian} (or {\it Lagrangian density}).  Note that a Lagrangian  $L$ can be also  considered as a  smooth real valued function on the infinite jet space $J^\infty(\bR^m; \bR^n)$. \par
We now recall that there is   a special class  of vector fields on $J^\infty(\bR^m; \bR^n)$, called {\it variational symmetries of $\cL$},  whose associated 1-parameter groups of (local) diffeomorphisms satisfy the following conditions (see \cite{Ol}, Ch. 5): 
 \begin{itemize}
 \item[a)]  they leave invariant the set of  maps  
 $$j^\infty(\s): \cU \subset \bR^m  \longrightarrow J^\infty(\bR^m; \bR^n)\ ,$$
  given by the jets  $j^\infty_x(\s)$ of  $\cC^\infty$  local maps $\s: \cU \subset\bR^m  \to \bR^n$; 
 \item[b)] they transform $L$ into other Lagrangians $L'$ that  differs from $L$ by terms that  give trivial contributions to    the  Euler-Lagrange equations. 
 \end{itemize}
 A vector field of this kind  is called {\it trivial variational symmetry} if it vanishes on the jets  $j^{\infty}_x(\s)$ of  the solutions  $\s: \cU \subset\bR^m  \to \bR^n$  to the Euler-Lagrange equations \eqref{1.1}. Two variational symmetries are said to be {\it equivalent} if they differ by a trivial one.  It is known that any equivalence class contains  a  subclass of  elements  $\vv_Q$ in special form, each of them uniquely determined by  a special $n$-tuple of differential operators $Q = (Q^1, \ldots Q^n)$.  Any such  element  is called {\it variational symmetry in evolutionary form}.\par
\smallskip
Olver's proof of Noether Theorem is crucially based on the following
\begin{theo} \label{OlverTheorem} Let $L$ be a Lagrangian  of order $k'$ and $F_\nu = 0$, $1 \leq \nu \leq n$, its associated  system of Euler-Lagrange equations of order $k = 2k'+1$. Suppose  also that for any $r$ its prolonged p.d.e. system $F^{(r-k)} = (F^{(r-k)}_\nu) = 0$  satisfies appropriate   constant  rank conditions. \par
Then a given (locally defined) $m$-tuple of   smooth  differential operators $P= (P^1, \ldots, P^m)$ of order $r$  satisfies a conservation law for  the Euler-Lagrange equations $F_\nu = 0$  if and only if 
it is equivalent (i.e. it differs by an $m$-tuple satisfying a trivial conservation law) to an $m$-tuple $\wt P$, whose divergence  $\Div \wt P$  has the form
\beq \label{1.5} \Div \wt P =  \sum_{\r = 1}^{n} Q^{\r} F_\r \eeq
 where  $Q = (Q^1, \ldots, Q^n)$ is the $n$-tuple  associated with a  variational symmetry $\vv_Q$ of $L$ in evolutionary form. 
 \end{theo}
We remark that the    constant  rank condition on $F^{(r-k)}$  is needed just  in the proof of the  ``only if'' part and that,  for {\it any} Euler-Lagrange equation,  a variational symmetry always determine  a conservation law.  \par
From this result the following general version  of Noether Theorem follows. \\[6pt]
\noindent{\bf Noether Theorem.}{ \it If $L$  is a Lagrangian having  prolongations of the associated Euler-Lagrange equations satisfying appropriate   conditions on ranks, local solvability and existence of non-characteristic directions (more precisely,  they are normal and totally nondegenerate systems; see  \cite{Ol} for definitions),  
then there exists   a one-to-one correspondence between  
\begin{itemize}
\item[a)] conservation laws for the Euler-Lagrange equations of $L$, determined up to additions of  trivial conservation laws; 
\item[b)] variational symmetries of $L$, determined up to  additions of trivial variational symmetries.
\end{itemize}
}
This version of Noether Theorem is based on  the   map between symmetries and conservation laws   determined by  \eqref{1.5}.  Note that such map goes from the space of variational symmetries  in evolutionary form  to the space of divergences, not  into  the space of the conserved $m$-tuples $P = (P^i)$. \par

\medskip
\subsection{An outline of our approach} 
\subsubsection{Holonomic submanifolds and holonomic distributions  on jet spaces}\hfill\par
Consider a bundle $\pi: E\longrightarrow M$ over an $m$-dimensional  oriented manifold $M$. Since all our discussions are of purely local nature, for simplicity, from now  on we assume that $M = \bR^m$,  oriented by  the standard volume form $\o = dx^1\wedge \ldots \wedge dx^m$. \par
For a given $k$-th order   jet space $\pi^k: J^k(E) \to \bR^m$,  any  (local) section $\s: \cU \subset \bR^m \longrightarrow E$ is uniquely associated with  the submanifold $(j^k\s)(\cU)$ of $J^k(E)$, given   by  their $k$-th order jets $j^k_x(\s)$, $x \in \cU$.  These submanifolds    are usually called {\it holonomic } (\cite{Gr}) and  can be characterised as  the only $m$-dimensional submanifolds of $J^k(E)$  
with:  
\begin{itemize}
\item[a)]  maximal rank  projections onto  $M$; 
\item[b)]  all tangent spaces  are contained in the vector spaces of   a special distribution $\cD \subset T J^k(E)$. 
\end{itemize}
Being related  with the holonomic sections, we call   such $\cD$ the {\it holonomic distribution of $J^k(E)$}. Note that  in other places such  distribution is called differently, as for instance   {\it canonical differential system}  (\cite{Ya, Ya1}) or   {\it Cartan distribution} (\cite{KLV, BCD}).  \par
A (locally defined)  $r$-form $\l$ on $J^k(E)$ is called {\it holonomic} if 
\begin{itemize}
\item[a)]  either $0 \leq r \leq m $ and    $\l$ vanishes when it is evaluated on $r$ vector fields in $\cD$ or
\item[b)]  $r \geq m+1$ and $\l$ vanishes when it is evaluated on at least $m$ vector fields in $\cD$.
\end{itemize}
If we set $s :=  \min\{r, m\}$,   we may also say that  an  $r$-form $\l$ is holonomic if and only if its restriction to an $m$-dimensional holonomic submanifold    vanishes identically when  it is evaluated on at least $s$ vector fields  that are tangent to  such submanifold.
\par
 \subsubsection{Lagrangians and actions}\hfill\par
We now observe that any functional  on the class of sections of $\pi: E \to \bR^m$ of the form 
$\cI = \int_{\cU} L(j^k_x(\s)) d x^1 \wedge \ldots \wedge d x^m$
  can be considered as a  functional on the class of (oriented) holonomic submanifolds of $J^k(E)$,  defined by 
  \beq \label{1.7}\cI\big|_{j^k\s(\cU)} \= \int_{j^k(\s)(\cU)} \a_L\ , \qquad \a_L \= L d x^1 \wedge \ldots \wedge d x^m\ .\eeq
Here we use the notation   $ \int_S \a_L$ to indicate the integral of the restriction of  $\a_L$ to the tangent space of $S$.   
\par
\medskip
The following fact is a crucial ingredient  of our construction (see \S \ref{sec2.3}):    in the class of  fixed boundary variations a holonomic submanifold $ j^k(\s)(\cU)$ is a  stationary point  for  $\cI$ if and only if it  is a  stationary point  for  any other  functional $\cI' = \int_{j^k(\s)(\cU)} (\a_L + \l + d \mu)$  with    
$\l$ and $\mu$  holonomic. Due to this,  we say that two $m$-forms $\a$, $\a'$ on $J^k(E)$ are  {\it variationally equivalent} if $\a - \a' = \l + d \mu$ for some  $\l$ and $\mu$  holonomic and we observe that  a variational principle  for   \eqref{1.7} can be considered as uniquely  associated with  the variational equivalence class $[\a_L]$ of  $\a_L$.  \par
\subsubsection{Conserved quantities as  differential forms}\hfill\par
   Consider an $m$-tuple of smooth $r$-th order differential operators  $P = (P^1, \ldots, P^m)$    and the  associated $(m-1)$-form  on $J^r(E)$   
  \beq\label{1.8} \h_P = \sum_{j = 1}^m (-1)^{m-1} P^j dx^1\wedge  \ldots \underset{j}{\wh{\phantom{j}}}\ldots \wedge d x^m\ .\eeq 
 Given a section  $\s: \cU \to E$, one can check  that $\Div P|_{j^k \s(\cU)} = 0$  if and only if the restriction $d \h_P|_{T (j^k\s(\cU))}$ of the  differential $d \h_P$ to the tangent space of 
$j^r \s(\cU) \subset J^r(E)$  is  identically equal to $0$. Further,   one has (see \S \ref{subsect41}):  
\begin{itemize}
\item[(1)]  $d \h_P|_{T (j^r\s(\cU))} = 0$  if and only if $d \h'|_{T (j^r\s(\cU))} = 0$ for 
any    $(m-1)$-form $\h' = \h_P+   \mu + d \nu$  with   $\mu$, $\nu$  holonomic; 
\item[(2)]  the integrals of $\h_P$ and $\h' = \h_P+   \mu + d \nu$
on  any closed $(m-1)$-dimensional submanifold of a holonomic submanifold  are equal. 
\end{itemize}
This  motivates the  following  definitions. We say that   two $(m-1)$-forms $\h$, $\h'$ on $J^r(E)$ are    {\it variationally equivalent} if $\h -  \h' =  \mu + d \nu$ for some holonomic forms $\mu$, $\nu$. Moreover,   given an  $m$-tuple of $r$-th order differential operators $P = (P^i)$, we call {\it variational class  of $P$} the equivalence class $[\h_P]$ of $(m-1)$-forms on $J^r(E)$  that are variationally equivalent to  $\h_P$.  \par
By (1) and (2),  $P$ satisfies a conservation law for a differential system if and only if the differential  of  an  $(m-1)$-form $\h$ in the variational class $ [\h_P]$  vanishes identically when  restricted to 
the  tangent spaces of  the
holonomic submanifolds associated with solutions. \par
\subsubsection{Infinitesimal $\cI$-symmetries and Noether Theorem}\hfill \par
\label{sect224}
Let  $L$ be a smooth Lagrangian  on $J^k(E)$ and $\cI$  the  functional  \eqref{1.7}  on holonomic submanifolds.  We call {\it weak (infinitesimal) symmetry for $\cI$} or, shortly,  {\it weak $\cI$-symmetry} any vector field on    $J^k(E)$   that generates a  1-parameter group of (local) diffeomorphisms  which  \par
\begin{itemize}
\item[(1)] preserve the holonomic distribution  $\cD$ or, more precisely, a slightly weaker condition, namely they map a special    subset of the vector fields in $\cD$  into vector fields in $\cD$ (see details in Definition \ref{defDsymmetries}), and
\item[(2)]  map an  element $\a \in [\a_L]$ into   $m$-forms  of  the same variational class. 
\end{itemize} 
Using coordinates,  one can    check that the vector fields   on $J^\infty(\bR^m, \bR^n)$ satisfying  (1) and (2)  coincide with  the vector fields that  
Olver calls  {\it variational symmetries}. Hence our weak $\cI$-symmetries can be considered as finite-dimensional versions  (defined  in a coordinate free language)  of Olver's variational symmetries. We also have to mention that even  the vector fields  that are called {\it Noether symmetries}   in \cite{BCD} are  related with our weak $\cI$-symmetries. In fact, using coordinates, one can  check that  they locally coincide with  Olver's {\it variational symmetries  in evolutionary form}. Hence, they correspond to  a  special subclass  of our  weak  $\cI$-symmetries.
\par 
\medskip
Our main result is the following (Theorems \ref{directNoether} and \ref{inverseNoether}). 
\begin{theo}\label{Theorem12}
Let $L:J^{k}(E) \to \bR$ be a smooth Lagrangian that  depends only on jets components of  order $k'$ satisfying the inequality $ 2 k' + 2 \leq \left[\frac{k}{2}\right]$. Then there exists an $m$-form $\a$ in the variational class of $\a_L = L dx^1\wedge \ldots \wedge dx^m$  with the following  properties. 
\begin{itemize}
\item[i)]   For any  (weak) $\cI$-symmetry $X$ on $J^k(E)$,  the $(m-1)$-form 
$\h = \imath_X \a$ is associated with an  $m$-tuple $P = (P^i)$ of $k$-th order differential operators  satisfying a conservation law for the Euler-Lagrange equations of $L$.
\item[ii)]   Let $k_o \leq  \left[\frac{k}{2}\right] -1$ and  $\cW \subset  J^k(E)$ be an open subset of the domain of $\a$ where the  Euler-Lagrange  equations $\cE(L) = 0$ of $L$ have a prolonged system with appropriate conditions on ranks and on the family  of jets of its solutions.  For any $m$-tuple of $k_o$-th order differential operators  $P = (P^i)$ on $\cW$, satisfying  a conservation law for  $\cE(L) = 0$, 
 there exists a weak  $\cI$-symmetry $X$ such that
$$\imath_X  \a = \h_P + \gz_{P'}$$
where $\h_P$ is defined  in \eqref{1.8} and $\gz_{P'}$ is  an $(m-1)$-form  corresponding to an $m$-tuple  $P' = (P'{}^i)$  satisfying  a trivial conservation law. 
\end{itemize}
 \end{theo}
As we mentioned in the Introduction, the $m$-form   $\a$  is called  {\it of Poincar\'e-Cartan type}  (see \ref{section45} for details) and it corresponds to the form $S \O$ defined by Kupershmidt  in \cite{Ku}, \S II.3.  The proof of   Prop. A2 in \cite{Sp} (see also \cite{Ge}, Thm.1.3.11)  provides an algorithm to   determine  an   $m$-form of  Poincar\'e-Cartan  type  for any  given  Lagrangian.\par
 \medskip
 \subsection{Comparisons with previous versions of Noether Theorem}\hfill\par
 \smallskip
The above Theorem \ref{Theorem12} yields  the existence of a one-to-one correspondence  between equivalence classes of weak  $\cI$-symmetries and equivalence classes of 
 conservation laws, exactly as it is implied by   Olver's Theorem \ref{OlverTheorem} or   other  versions of Noether Theorem (see  e.g.  \cite{BCD}, \S 5.4.1).
On the other hand, in our approach  such  correspondence is   determined by means of  a very  simple linear   map, namely   the contraction map 
 $X \mapsto  \imath_X \a$ with an $m$-form $\a$ of Poincare-Cartan type. This  gives a  direct way to go
  from the    weak $\cI$-symmetries  of $k$-th order  into   conserved  $m$-tuples  $P$ of $k_o$-th order operators,   {\it not   into the space of  divergence operators}    as it occurs  in Olver's   and other versions of  Noether Theorem. Further, this map is surjective, in the sense that {\it any} conserved $m$-tuple $P$ is, modulo addition of $m$-tuples satisfying trivial conservation laws, is in the image of the above described linear map. 
\par
\smallskip
Another result of our  approach is  the    unveiling   of the importance of a  distinguished  relation between the  Poincar\'e-Cartan $1$-form of Hamiltonian Mechanics  and conservation laws, a relation that  generalises   to {\it all}  smooth systems of  ordinary and  partial differential equations of variational origin.   
We also point out that  all notions considered in our construction   are expressed in terms of  standard differential geometric objects.  The   proofs use only basic  properties of differential forms on finite-dimensional manifolds, as for instance  Stokes' Theorem and   Homotopy Formula. This  paves the way  to  direct extensions of Noether Theorem   to  many other interesting settings, as e.g. to   supergeometric contexts. We  plan  to undertake this task in  future papers.
\par
\smallskip
We conclude recalling   that  a direct  correspondence between symmetries and conserved quantities  was also  established  by Lychagin    for the Euler-Lagrange equations that are in the class of   {\it Monge-Amp\`ere equations}. This is  a large and important family  of  non-linear  second order differential equations  on real functions $f: \cU \subset \bR^m \to \bR$ of $m$ independent variables (see  \cite{Ly, KLR} and references therein). They are equations  usually denoted by  $\D_\o(f) = 0$ and they are equivalent to the vanishing of some fixed $k$-form $\o$ on $ J^1(E)$, $E = \bR \times \bR^m$,  on the  holonomic submanifold $j^1(f)(\cU)$ of the unknown function $f: \cU \subset \bR^m \to \bR$.  In the cases in which  $\D_\o(f) = 0$  coincides with  an  Euler-Lagrange equation,  Lychagin constructed    an explicit  linear map from the class of  symmetries of the  equation into the class of conserved quantities, which establishes the bijection of  Noether Theorem  (\cite{Ly}, Thm. 4.4).  We  expect that  Lychagin's  map  coincides with our map $X \mapsto \imath_X \a$ for  an appropriate choice of an  $m$-form $\a$ of  Poincar\'e-Cartan type.  \par
We observe  that   Lychagin's map can be constructed for {\it all Monge-Amp\`ere equations  of divergence type}, not only for those  of variational origin.   We expect  that  a deeper understanding of the relation between  $m$-forms of  Poincar\'e-Cartan type and Lychagin's map would   lead to interesting generalisations of Noether Theorem.     
 \par
\bigskip
\section{A differential-geometric presentation 
of  variational principles}
\label{geompresent}
\setcounter{equation}{0}
 \subsection{Notational remarks}
 \hfill\par
In what  follows, we consider  only  partial differential equations  on $\cC^\infty$ maps  from open subsets of $\bR^m$, oriented by  the  standard volume form $dx^1\wedge \ldots\wedge dx^m$,   into a  fixed   $n$-dimensional manifold $M$. Since any  such  map  $f: \cU\subset \bR^m \longrightarrow M$ is  uniquely determined by the associated    (local) section of the trivial bundle
$\pi: E = \bR^m \times M  \longrightarrow  \bR^m$
$$\s^{(f)}(x^1,\dots,x^m) \= (x^1,\dots, x^m,f(x^1,\dots,x^m))\ ,$$
 we always consider a system of partial differential equations  as a set of differential equations  on the   smooth sections  of the  bundle $ E$.   \par
\medskip
Given an integer   $k\geq 1$ and   a smooth section $\s: \cU \subset \bR^m \longrightarrow E = \bR^m \times M $, we  denote by  $j^k_{p}(\s)$ for the $k$-th order jet of $\s$ at $p \in \cU$. The space of all   $k$-jets  is denoted by $J^k(E)$.
For any $1 \leq \ell \leq k$, we set
$$\pi^k_\ell: J^k(E) \longrightarrow J^\ell(E)\ ,\qquad \pi^k_\ell(j^k_{p}(\s)) \= j^\ell_{p}(\s)$$
and we denote by  $\pi^k_0:J^k(E) \longrightarrow E$ and $\pi^k_{-1}: J^k(E) \longrightarrow \bR^m$  the  natural projections onto $E$ and $\bR^m$, i.e. the maps 
$$\pi^k_0(j^k_{p}(\s)) \= \s(p)\qquad \text{and}\qquad \pi^k_{-1}(j^k_{p}(\s)) \= p\ , \ \ \text{respectively}\ .$$
Given a  section $\s : \cU \subset \bR^m \longrightarrow E$,  we call   {\it  $k$-th order  lift} of $\s$  the map
$$\s^{(k)}: \cU \subset \bR^m \longrightarrow J^k(E)\ ,\qquad \s^{(k)}(p) \= j^k_p(\s)\ .$$
\par
Finally,   given a system of coordinates $\xi = (y^i): \cW \subset M \longrightarrow \bR^n$ on an open set $\cW \subset M$, we denote by $\wh \xi$  the  associated coordinates
$$\wh \xi: \bR^m \times \cW  \longrightarrow  \bR^{n+m}\ ,\qquad \wh \xi(p, q) := (x^1(p),\dots, x^m(p), y^1(q), \ldots, y^n(q))\ ,$$
where the $x^i$'s are  the standard coordinates of $\bR^m$.  The coordinates  $\wh \xi = (x^i, y^j)$ are called   {\it  associated with the coordinates $\xi = (y^i)$}. Any set of
coordinates   constructed  in this fashion  is called   {\it set of adapted coordinates}.\par
\smallskip
For a  given set of   adapted coordinates $(x^i, y^j)$,  we may consider the naturally  associated   set of  coordinates
\begin{multline}\wh \xi^{(k)} =  \left(x^i, y^j, (y^j_I)_{|I| = 1, \ldots, k}\right)  :  \cU \subset J^k(E) \longrightarrow \bR^{m +n + N}\ ,\\ N :=n \sum_{\ell=1}^k\binom{m+\ell-1}{\ell}\ ,
\end{multline}
defined  for any  $u = j^k_p(\s)$ in $ \cU = (\pi^k_0)^{-1}(\bR^m \times \cW)$ as follows:
\begin{itemize}
\item[a)] the coordinates $x^i (u) $, $1 \leq i \leq m$,   are the standard coordinates of $p = \pi^k_{-1}(u) \in \bR^m$;
\item[b)]  the coordinates $y^j  (u) $, $1 \leq i \leq  n$,  are the last $n$ coordinates of the set of adapted coordinates of   $(p, s(p)) = \pi^k_0(u) \in \bR^m \times M$;
\item[c)] the coordinates $y^j_I  (u) $,  with $1 \leq j \leq m$ and $I = (I_1, \ldots, I_m)$  multiindex of  order  $ |I| \= \sum_{j= 1}^m I_j$ with $1 \leq |I| \leq k$,  are the values of the partial derivatives
$$y^j_I  (u) \= {\left.\frac{\p^{|I|}  \s^j}{\p x^I}\right|}_{(x^1(p),\dots,x^m(p))}$$
of a section $\s$  in the equivalence class $u= j^k_p(\s)$.
\end{itemize}
The coordinates $\wh \xi^{(k)}$ are  called   {\it  adapted coordinates on $J^k(E)$ associated with the coordinates $\xi = (y^i)$}.\par
\medskip
\subsection{Holonomic  \texorpdfstring{$p$}\ -forms and variational classes} \label{section22}\hfill\par
\begin{definition}  The {\it holonomic submanifolds of $J^k(E)$} (see e.g. \cite{Gr})  are the submanifolds $\cS \subset J^k(E)$ for which there exists a section $\s: \cU \subset \bR^m \longrightarrow E$
such that $\cS = \{\ u \in J^k(E)\ :\ u = \s^{(k)}(x)\ ,\ x \in \cU\ \}$.\par
We call  {\it holonomic distribution of $J^k(E)$}  the distribution $\cD \subset T J^k(E)$ generated at any   $u \in J^k(E)$ by the vectors that are tangent to holonomic submanifolds, i.e., 
$$\cD_u = \Span\left\{\ v \in T_u J^k E\ :\ v =\s^{(k)}_*(w) \text{ for some}\ \ w\in T_p \bR^m\ \right.\phantom{aaaaaaaaaa}$$
$$\phantom{aaaaaaaaaaaaaaaaaaaaaaaaaaaa} \left.\text{and some}\ \ \s\ \  \text{such that}\ j^k_{p}(\s) = u   \right\}\ .$$
The vectors  in $\cD$ and the vector   fields with values in $\cD$  are called {\it holonomic} (see   \S 2 for other names often used  for the distribution $\cD$). 
\end{definition}
 Let  $\wh \xi^{(k)} = (x^i, y^j, y^j_I)$ be a set of adapted coordinates on some open set $\cU \subset J^k(E)$ and fix a jet  $\bar u = j^k_{p}(\bar \s) $  with coordinates  $\wh \xi^{(k)} (\bar u) = (\overline x^i,  \overline y^j, \overline y^j_I)$. The  vectors $v \in T_{\bar u} J^k(E)$ having the form  $v=\s^{(k)}_*(w)$ for some  $w=w^i\left.\frac{\partial}{\partial x^i}\right|_{p} \in T_{p}\bR^m$  and some section $\s$  with  $j^k_p(\s) = \bar u$,  are 
\begin{multline} \label{strictlyhol}  v= w^i \left(\left.\frac{\partial}{\partial x^i}\right|_{(\overline x^i,  \overline y^j, \overline y^j_I)} + \sum_{0 \leq|I| \leq k-1} \overline y^j_{I + 1_i} \left.\frac{\partial}{\partial y^j_I} \right|_{(\overline x^i,  \overline y^j, \overline y^j_I)}+\right.\\
+\left.\sum_{|J| = k} \left.\frac{\partial^{|J| + 1} \s^j}{\partial x^{J+ 1_i}}\right|_{\overline x}\left.\frac{\partial}{\partial y^j_J} \right|_{(\overline x^i,  \overline y^j, \overline y^j_I)}\right)\ .\end{multline}
(here, given $J = (J_1, \ldots, J_m)$, we set $J + 1_i\= (J_1, \ldots, J_i +1, \ldots J_m)$).\par
Since  the values $\left.\frac{\partial^{|J| + 1} \s^j}{\partial x^{J+ 1_i}}\right|_{\overline x}$, $|J| = k$,  may vary arbitrarily   by  making different choices for  $\s$  in  the  $k$-th order jet   $\bar u = j^k_p(\s)$,  we have that      $\cD_u \subset T_u J^k(E)$  {\it is  generated by the linearly independent  vectors}
\begin{multline} \left.\frac{d}{d x^i} \right|_{(\overline x^i,  \overline y^j, \overline y^j_I)}\= \left.\left(\frac{\partial}{\partial x^i} + \sum_{0 \leq|I| \leq k-1} \overline y^j_{I + 1_i} \frac{\partial}{\partial y^j_I}\right) \right|_{(\overline x^i,  \overline y^j, \overline y^j_I)}\\ \text{and}\qquad
 \left.\frac{\partial}{\partial y^j_J}\right|_{(\overline x^i,  \overline y^j, \overline y^j_I)}\ \text{with}\ \ |J| = k\ .\end{multline}
 \par
 \smallskip
The notion of holonomic distribution  leads to  the following.
\begin{definition}
A (local) $p$-form $\l$ of $J^k(E)$   is called  {\it holonomic} if it satisfies one of  the following conditions:
\begin{itemize}
\item[a)] $p\leq m$ and   for any  $p$-tuple $(X_1, \ldots, X_p)$ of  holonomic vector fields, one has  $\l(X_1,\dots,X_p)=0$;
\item[b)]  $p>m$ and  for any  $m$-tuple $(X_1, \ldots, X_m)$ of holonomic vector fields,  one has $\l(X_1,\dots,X_m,\star,\dots,\star)=0$.
\end{itemize}
If $\a$, $\a'$ are $p$-forms  on the same open subset $\cU \subset J^k(E)$,
 we call them  {\it variationally equivalent} if there exist a holonomic $p$-form  $\l$ and a holonomic $(p-1)$-form  $\mu$ such that
$$\a' =\a +  \l + d \mu \ .$$
For a fixed  $ \cU \subset J^k(E)$,
the  variational equivalence is an equivalence relation  on the set of    $p$-forms   on $\cU$.  The equivalence class of  $\a$  is called {\it variational class of $\a$} and is denoted by  $[\a]$.\par
Finally, we say that a  $p$-form $\a$ is  {\it proper} if $\imath_V \a = 0$ for any  vector field $V$ that  is  vertical  with respect to the projection  $\pi^k_{k-1*}: TJ^k(E) \to T J^{k-1}(E)$. \par
\end{definition}
The  role played by holonomic forms and variational classes in our approach has been  shortly described  in \S \ref{Olver's proof}.  See    \S \ref{sec2.3} below  for  further details.  \par
\smallskip
 The   explicit expressions in coordinates
of   holonomic $q$-forms is quite helpful to get a better understanding of these objects.  To write them down,  we first need to impose  the following order on the set of  indices:
\begin{itemize}
 \item[a)] the  multiindices  are  subjected to the   lexicographic order, namely given    $J=(J_1,\dots, J_m)$ and  $J'=(J'_1,\dots,J'_{m})$, we say that $J<J'$ if and only if $|J| < |J'| $  or $ |J| = |J'|$ and  there exists $\ell\leq m$ such that  $j_i=j'_i$ for $i = 1, \ldots,  \ell -1$ and  $j_{\ell}<j'_{\ell}$;
\item[b)]  given two pairs $(j,J)$ and $(j',J')$ with $1 \leq j,j' \leq n$ and $J$, $J'$ multiindices,  we write  $(j,J)<(j',J')$ to indicate that either  $J<J'$ or $J=J'$ and $j<j'$.
\end{itemize}
Consider now the collection of $1$-forms
\beq  \label{holonomicforms}  \begin{array}{ll} dx^i&\mbox{for}\quad 1\leq i\leq m\ ,\\
\ & \\
\omega^j_J\=dy^j_J-\sum_{\ell =1}^my^j_{J+1_\ell}dx^\ell\ &\mbox{for}\quad 1\leq j\leq n\ ,\ 0\leq |J|\leq k-1\ ,\\
\ & \\
\psi^j_L\=dy^j_L & \mbox{for}\quad 1\leq j\leq n\ ,\ |L|=k\ .
\end{array}\eeq
This collection of $1$-forms  gives a basis for $T^*_u J^k(E)$ at any $u$, so that any  $q$-form $\a$ can be written as a linear combination of wedge products of such $1$-forms and it can be written as
\beq \label{q-forma}\sum_{\begin{subarray}{c} \ell +r+s=q\\
i_1<\ldots <i_\ell\\
(j_1,J_1)<\ldots <(j_r,J_r)\\
(k_1,L_1)<\ldots <(k_s,L_s)\\
0 \leq |J_s|\leq k-1\ ,\ |L_j|=k
\end{subarray}}
\hskip - 1cm \a_{i_1\dots i_\ell|j_1\dots j_r|k_1\dots k_s}^{\phantom{i_1\dots i_\ell}J_1\dots J_r|L_1\dots L_s}
dx^{i_1}\wedge \ldots \wedge dx^{i_\ell}\wedge \o^{j_1}_{J_1}\wedge \ldots \wedge \o^{j_r}_{J_r}\wedge \psi^{k_1}_{L_1}\wedge \ldots \wedge \psi^{k_s}_{L_s}\ .\eeq
From this expression and the definition of $\psi^j_J$, we see that  $\a$ is  proper if and only if it is of the form
\beq\label{q-formapropria}\a= \sum_{\begin{subarray}{c} \ell +r=q\\
i_1<\ldots <i_\ell\\
(j_1,J_1)<\ldots <(j_r,J_r)\\
0 \leq |J_s|\leq k-1\ ,
\end{subarray}}
\!\!\!\!\!\!\!\!\!\!\!\!\!\!\!\a_{i_1\dots i_\ell|j_1\dots j_r}^{\phantom{i_1\dots i_\ell}J_1\dots J_r}
dx^{i_1}\wedge \ldots \wedge dx^{i_\ell}\wedge \o^{j_1}_{J_1}\wedge \ldots \wedge \o^{j_r}_{J_r}\ .
\eeq
On the other hand, $\a$  is holonomic if and only if it is determined by an expression \eqref{q-forma} satisfying one of these conditions:
\begin{itemize}
\item[a')] $q < m$ and  all coefficients of  terms with $\ell + s = q$ (hence, with $r=0$) are equal to $0$;
\item[b')]  $q \geq m$ and  all coefficients of  terms with $\ell +s \geq m$ (hence, with $r\leq q-m$) are equal to $0$.\end{itemize}
Consequently  a {\it proper}  $q$-form $\a$ is holonomic if and only if it admits one  of the following expressions:
\begin{itemize}
\item[a'')] $q<m$ and
$$\a = \sum_{\!\!\!\begin{subarray}{c}\ell +r=q
\\  1 \leq r \leq q\end{subarray}}\ \sum_{\begin{subarray}{c}
i_1<\ldots <i_\ell\\
(j_1,J_1)<\ldots <(j_r,J_r)\\
0 \leq |J_s|\leq k-1\ ,
\end{subarray}}\!\!\!\!\!\!\a_{i_1\dots i_\ell|j_1\dots j_r}^{\phantom{i_1\dots i_\ell}J_1\dots J_r}
dx^{i_1}\wedge \ldots \wedge dx^{i_\ell}\wedge \o^{j_1}_{J_1}\wedge \ldots \wedge \o^{j_r}_{J_r}\ ;
$$
\item[b'')] $q\geq m$ and
$$\a = \sum_{\!\!\!\begin{subarray}{c}\ell +r=q
\\  q-m+1 \leq r\leq q\end{subarray}}\ \sum_{\begin{subarray}{c}
i_1<\ldots <i_\ell\\
(j_1,J_1)<\ldots <(j_r,J_r)\\
0 \leq |J_s|\leq k-1\ ,
\end{subarray}}
\!\!\!\!\!\!\a_{i_1\dots i_\ell|j_1\dots j_r}^{\phantom{i_1\dots i_\ell}J_1\dots J_r}
dx^{i_1}\wedge \ldots \wedge dx^{i_\ell}\wedge \o^{j_1}_{J_1}\wedge \ldots \wedge \o^{j_r}_{J_r}\ .
$$
\end{itemize}
\medskip
These formulae are quite useful to quickly check several properties of holonomic forms. For instance, one can directly see that {\it  the differential $d \a$ of a  holonomic form $\a$  needs not   be holonomic}.\par
\bigskip
\subsection{Variational classes, Lagrangians and source forms}
\label{sec2.3}
\hfill\par
We now consider   variational principles for   functionals   of the form
\beq\label{action1} I_L(\s) = \int_{\cU} \big(L\circ \s^{(k)}\big)(x^1, \ldots, x^m) dx^1\wedge \ldots \wedge dx^m\ ,  \eeq
 determined by a smooth Lagrangian $L: J^k(E) \longrightarrow \bR$.  As it was explained in   \cite{Sp} (see also \cite{Ge, FS}),    the functionals \eqref{action1}  can be considered  as special  cases of a slightly  larger class of  functionals, which we now recall.
\par
\begin{definition} \label{defaction}  Let  $[\a]$ be a variational class of $m$-forms  in $J^k(E)$. We call {\it action  determined by  $[\a]$}
the functional on  sections $ \s:\cU\subset \bR^m\longrightarrow E$  on regions $\cU$ with  piecewise smooth boundary $\partial \cU$,    defined by
\beq \label{action2}  \cI_{[\a]}(\s)\=\int_{\s^{(k)}(\cU)}\a\ .\eeq
Here, we use the  notation   $\int_{\s^{(k)}(\cU)}\a$ to indicate   the integral  of  an $m$-form $\a$ in the variational  class $[\a]$, restricted  to the points and to  the tangent vectors of the oriented $m$-dimensional submanifold  $\s^{(k)}(\cU) $ of $J^k(E)$.
\end{definition}
\par
\medskip
We stress the fact  that,   by the  very definition of variational classes, {\it the integral $\int_{\s^{(k)}(\cU)}\a$ is independent on the choice of the representative $\a$ in $[\a]$} and it is therefore well defined. Indeed,  if $\a$, $\a'$ are variational equivalent, i.e.   $\a' = \a + \l + d \mu$ for some holonomic  $m$-form $\l$  and holonomic $(m-1)$-form $\mu$, by Stokes' Theorem and the fact that the vectors that are tangent to  $\s^{(k)}(\cU)$ are holonomic,
$$\int_{\s^{(k)}(\cU)}\a' = \int_{\s^{(k)}(\cU)}\a + \int_{\s^{(k)}(\cU)}\l +  \int_{\partial \s^{(k)}(\cU)}\mu= \int_{\s^{(k)}(\cU)}\a\ .$$
Furthermore, {\it any functional of the form \eqref{action1} can be considered as an action of the form \eqref{action2}}. Indeed,
if  $L: J^k(E) \longrightarrow \bR$ is a function of class $\cC^\infty$  and if we set $\a_L \= L\, \pi^k_{-1}{}^*(dx^1\wedge \ldots \wedge dx^m)$,  we  see that for any section $\s$
$$\cI_{[\a_L]} (\s)= \int_{\s^{(k)}(\cU)}\a_L = \int_\cU (L\circ \s^{(k)})dx^1\wedge \ldots \wedge dx^m = I_L(\s) \ . $$
Conversely,   {\it any action having the form \eqref{action2} can be locally identified with a functional of the form \eqref{action1}}. To see this,   let $\a$ be an $m$-form on $J^{k}(E)$ and consider the pull-back  $\wt \a \= (\pi^{k+1}_k)^*(\a)$ on  $J^{k+1}(E)$. Being $\wt \a$ proper, its  expression  in    adapted coordinates $\wh \xi^{(k)} = (x^i, y^j, y^j_{I})$  has the form (see \eqref{q-formapropria}):
\begin{multline}\label{q-formapropria-bis}
\wt \a= \a_0 dx^1 \wedge \ldots \wedge d x^m +\\
+ \displaystyle
\sum_{\!\!\!\ell=1}^{m-1}\bigg(\!\!\!\sum_{\begin{subarray}{c}
i_1<\ldots <i_\ell\\
(j_1,J_1)<\ldots <(j_{m-\ell},J_{m-\ell})\\
0 \leq |J_s|\leq k-1\ ,
\end{subarray}}
\!\!\!\!\!\!\!\!\!\!\!\!\!\!\!\a_{i_1\dots i_\ell|j_1\dots j_{m-\ell}}^{\phantom{i_1\dots i_\ell}J_1\dots J_{m-\ell}}
dx^{i_1}{\wedge} \ldots {\wedge} dx^{i_\ell}{\wedge} \o^{j_1}_{J_1}{\wedge} \ldots {\wedge} \o^{j_{m-\ell}}_{J_{m - \ell}}\bigg).
\end{multline}
We now observe that here  all terms except the first one  are holonomic. Hence  $[\wt \a] = [\a_0 dx^1 \wedge \ldots \wedge dx^m]$. This means that the values of    $\cI_{[\a]} $ on  sections with values in the domain $\cW$ of the adapted coordinates coincide  with those  given by  the functional   \eqref{action1} with    $L \= \a_0|_{\cW}$. \par
\medskip
These  remarks show  that {\it the class of functionals introduced with  Definition \ref{defaction}  is a natural extension  of the class of usual   actions  \eqref{action1}}. \par
\medskip
We conclude   introducing the following convenient  terminology . Let   $\wt \b$ be a  (locally defined)  $p$-form   on a jet space $J^k(E)$. We say that $\wt \b$   is   {\it of order $r$}  if  we can write $\wt \b = (\pi^k_r)^* \b$ for some   (locally defined) $p$-form $\b$  on $J^r(E)$, $0\leq r \leq k$.
 According to this definition,  {\it any $p$-form $\b$  on a jet space  $J^r(E)$  can be naturally identified  with a $p$-form  of order $r$  on any other jet space $J^k(E)$ with $k \geq r+1$}. Further, note that  {\it if a   $p$-form  $\b'$ on $J^{k}(E)$ is    of order $0 \leq r \leq k-1$, then it  is     proper}. \par
Due to this  it is  possible to  identify any (not necessarily proper)   $p$-form on a jet space $J^r(E)$   with  a {\it proper}  $p$-form (of order $r$) on a jet space $J^k(E)$ with  $k \geq r+1$. This shows that, in many arguments,  there is no loss of generality if  one reduces to consider  only proper $q$-forms. \par
\bigskip
\subsection{Variational Principles  and Euler-Lagrange equations}\hfill\par
\label{section2.4}
We now   consider    variational principles  for the actions defined  in Definition \ref{defaction}.   As the reader will shortly see, our presentation  is   designed
to derive  from   a given variational principle the same   Euler-Lagrange equations that one obtains from  Lagrangians    in  usual settings. 
\par
Consider a section   $\s: \cU \longrightarrow E$  and a regular $m$-dimensional region $D$ in $\cU \subset \bR^m$. With the expression ``{\rm regular region}'' we mean a  
 connected   open subset $D \subset \cU$,  whose closure $\overline D$ is an $m$-dimensional oriented manifold  with corners (see e.g.  \cite{Jo}  for the definition).\par
A smooth  map $F: D \times (-\ve, \ve)\longrightarrow E$ is called {\it   variation  of $\s|_D$  with fixed $k$-th order boundary}  if it    satisfies  the following  conditions:
\begin{itemize}
\item[a)]  the  maps $F^{(s)} \= F(\cdot, s): D \longrightarrow E$,  $s \in (-\ve, \ve)$, are such that $F^{(0)}  = \s$ and, for any $s$, the map  $F^{(s)}$   is smoothly extendible   to  $\overline D$;
\item[b)]  for any $s\in (-\e,\e)$,   the $k$-order lift $(F^{(s)})^{(k)}  \= j^{k}{(F^{(s)})}$ of the extension $F^{(s)} : \overline D \to E$ satisfies the boundary condition
${(F^{(s)})^{(k)}}\big|_{\partial D}=\s^{(k)}\big|_{\p D}$.
\end{itemize}
\begin{definition}\label{variational}
Let  $[\a]$ be a variational class of $m$-forms on $J^k(E)$ and  $\s:\cU\subset \bR^m\longrightarrow  E$ a section. We say that $\s$  {\it satisfies the variational principle  of  $\cI_{[\a]}$} if for any regular region $D \subset \cU$ and any variation $F$ of $\s|_D$  with fixed $k$-th order boundary, one has
\beq \label{stationarity} \left.\frac{d \big(\cI_{[\a]}(F^{(s)})\big)}{ds}\right|_{s=0}=\frac{d}{ds}\big(\int_{j^k(F^{(s)})(D)}\a\big)\bigg|_{s=0}=0\ .\eeq
\end{definition}
\par
\bigskip
We now want to show that  the sections that satisfy such variational principle are precisely the  solutions  to  the usual  Euler-Lagrange equations of  classical setting. For this, we first need to   reformulate  \eqref{stationarity} into an equivalent condition   involving  a special kind of  vector fields. \par
\medskip
 Let $\s: \cU \longrightarrow E$ be a section, $D \subset \cU$ a regular region and
 $W: \s^{(k)}(\overline D)\longrightarrow T J^k(E)|_{\s^{(k)}(\overline D)}$
 a vector field   defined only   at the points  of $\s^{(k)}(\overline D)$.  We say that  $W$ is  a {\it $k$-th order variational field}  if there exists a smooth variation $F: D\times (-\ve,\ve)\longrightarrow E$ of $\s$   with fixed $k$-th order boundary  such that
 \beq \label{variationalfield} W= F^{(k)}_*\left({\left.\frac{\p}{\p s}\right|}_{(x,0)}\right)\ ,\qquad F^{(k)}(x,s)\=j^k_x(F(\cdot,s))\ . \eeq
 We remark  that, by the property (b) of the variations with fixed $k$-th order boundary, any variational vector  field $W$  is such that  
 \beq \label{vanishing}ÊW|_{\s^{(k)}(\p D)} = 0\ .\eeq
\par
\medskip
\begin{prop}\label{TeoGusteaux}
A section $\s:\cU\longrightarrow E$ satisfies the variational principle of  $\cI_{[\a]}$ if and only if
\beq \label{infinitesimalstationarity} \int_{ \s^{(k)}(D)}\imath_W d \a = 0\eeq
for any regular region $D \subset \cU$ and any $k$-th order variational field $W$  on $\s^{(k)}(\overline D)$.
\end{prop}
\begin{pf}  Let  $W$ be the variational field  \eqref{variationalfield} determined by  a smooth variation  $F$ with fixed $k$-th order boundary.
By   Stokes' Theorem for manifolds with corners (see e.g. \cite{Le}) and by \eqref{vanishing},   for any $m$-form $\a$ on $J^k(E)$ and any $h \in (- \ve,  \ve)$
$$\int_{(F^{(h)})^{(k)}(D)} \a - \int_{(F^{(0)})^{(k)}(D)} \a = \int_{D \times \{h\}} F^{(k)}{}^* (\a) - \int_{D \times \{0\}} F^{(k)}{}^* (\a) = $$
$$ =  \int_{\p(D \times (0,h))} F^{(k)}{}^* (\a) =  \int_{D \times (0,h)} F^{(k)}{}^* (d \a) \ .$$
Hence, 
\begin{align*} \left.\frac{d \big(\cI_{[\a]}(F^{(s)})\big)}{ds}\right|_{s=0} &= \lim_{h \to 0} \frac{1}{h} \left(\int_{(F^{(h)})^{(k)}(D)} \a - \int_{(F^{(0)})^{(k)}(D)} \a \right) =\\
&= \lim_{h \to 0} \frac{1}{h}  \int_{D \times (0,h)} F^{(k)}{}^* (d \a)  =    \int_{D} \s^{(k)}{}^* \big((-1)^m \imath_ W d \a\big) = \\
& = 
(-1)^m \int_{\s^{(k)}(D)} \imath_W d \a\ .
\end{align*}
From this the  claim  follows.
\end{pf}
 In absence of  an effective   characterisation  of the  variational vector fields,
 condition  \eqref{infinitesimalstationarity}  does not seem to correspond to  any  system  of partial differential equations for  $\s$.
On the other hand,  we have to stress   that  if $\s$  satisfies  \eqref{infinitesimalstationarity}  for a given choice of  variational vector field  $W$, it also  satisfies   the equality $ \int_{\s^{(k)}(D)}\imath_W \b = 0$  for any $(m+1)$-form $\b$ which is variationally equivalent to  $d\a$. Indeed, if $\b= d\a + \l + d\mu$   for some holonomic $\l$ and $\mu$, by   Stokes' Theorem,   holonomicity and \eqref{vanishing}, we have
\beq
\begin{split} \label{infinitesimalstationarity1}  \int_{\s^{(k)}(D)}\imath_W d \a &
= \int_{\s^{(k)}(D)}\imath_W \b -
\int_{\s^{(k)}(D)}\imath_W d\mu =\\
&= \int_{\s^{(k)}(D)}\imath_W \b -
\int_{\s^{(k)}(D)}\cL_W \mu + \int_{\s^{(k)}(D)}d(\imath_W \mu)= \\
& 
=  \int_{\s^{(k)}(D)}\imath_W \b -
\int_{\s^{(k)}(D)}\cL_W \mu  + \int_{\s^{(k)}(\p D)} \imath_W \mu =\\
& =  \int_{\s^{(k)}(D)}\imath_W \b -
\int_{\s^{(k)}(D)}\cL_W \mu \ .
\end{split}
\eeq
Here, $\cL_W \mu$ is to be understood as the Lie derivative of $\mu$ along  some  smooth extension of $W$ on a neighbourhood of $\s^{(k)}(\overline D)$. By definition of $W$, we may always assume that such local extension has a local flow $\Phi^W_t$, which  is the lift to $J^k(E)$ of a fiber preserving flow $\Phi^{\wt W}_t$ on $E$, generated by a local vector field $\wt W$ of $E$ that projects trivially on $\bR^m$. Under this assumption, the local flow $\Phi^W_t$ maps holonomic sections into holonomic sections, hence it preserves the holonomic distribution $\cD$. This yields that   the Lie derivatives of holonomic forms  by  $W$ are   holonomic and  that 
$  \int_{\s^{(k)}(D)}\imath_W d \a =
  \int_{\s^{(k)}(D)}\imath_W \b$, as claimed.\par
  \medskip
This fact motivates the importance of  some  special   representatives  of $[d \a]$, called {\it source forms} and which are now about to define. For this we  need to introduce   a  preliminary  notion  (see e.g. \cite{Sp, Ge}): A proper $q$-form $\b$ on a jet space  $J^k(E)$, $k \geq 1$,   is called
{\it homogeneous} if there are  non-negative integers $\ell$, $r$ such that  $\ell + r = q$  and so that,  for any set $\{X_1, \ldots, X_q\}$  of  $q$ vector fields  that contains either more than $\ell$  holonomic vector fields or more than $r$ vector fields   projecting trivially on  $\bR^m$, one has
$$\b(X_1,\ldots, X_{q}) = 0\ .$$
 If $\b$ is homogeneous and satisfies the above condition for the integers $\ell$ and $r$,  we call   the pair $(\ell, r)$ the  {\it bi-degree of} $\b$.
 It can be checked that the bidegree of a non-trivial proper homogeneous $q$-form $\b$ is uniquely associated with  $\b$. \par
\smallskip
\begin{definition} \label{defsource}
A {\it source form}  on  $J^k(E)$ is any (locally defined) $(m+1)$-form $\b$ which is proper, 
 homogeneous of bi-degree $(m,1)$ and such that
\beq \label{sourceform}  \b(X_1, \ldots, X_m, V)  = 0\eeq
for any   holonomic vector  fields $X_i$ and  any $\pi^k_0$-vertical vector field  $V$ (i.e.,   such that 
$\pi^k_{0*}(V) = 0$).
\end{definition}
For a better understanding of  source forms, it is convenient to see what are the coordinate expressions of these $(m+1)$-forms
in a  system of adapted coordinates $\wh \xi^{(k)} = (x^i, y^j, y^j_I)$.  One can directly check that   an $(m+1)$-form $\b$ is a source form if and only if   it has the form
\beq \label{sorgente}\b =\sum_{j =1}^n \b_j dx^1 \wedge \dots \wedge dx^m\wedge dy^j  =\sum_{j =1}^n \b_j dx^1 \wedge \dots \wedge dx^m\wedge \o^j_0\eeq
at all points where the coordinates are defined.
We also remark that  by Prop. A.2 in  \cite{Sp} (see also \cite{Ta, Ge, FS}) {\it given an $m$-form  $\a = L dx^1\wedge \ldots \wedge dx^m$   determined by a Lagrangian $L$  of order $r$,  the variational class   $[d\a]$ on a jet space $J^k(E)$ with $k \geq 2r$ contains  exactly one  source form $\b$.  Locally, such source form is given by the   coordinate expression \eqref{sorgente} in which the  
 components  $\b_j$ are determined  by  applying the classical  Euler-Lagrange operator  to $L$.  In particular,
when $r = 1$ and $k = 2$,   the explicit expressions  of the components $\b_j$ are }
 \beq \label{2.12}  \b_j \=  -\frac{\p L}{\p y^j}+\sum_{\ell = 1}^m\frac{d}{dx^\ell}\bigg(\frac{\p L}{\p y^j_\ell}\bigg) \eeq
 (for  properties of higher order Lagrangians and Euler-Lagrange operators, see e.g. \cite{An}, \S II.B).
 \par
\medskip
We are now able to show   that  a  section satisfies a variational principle for   $\cI_{[\a]}$ if and only if it satisfies the corresponding
Euler-Lagrange equations.
\begin{theorem}\label{cor2.3.4} Let $\a = L dx^1\wedge \ldots \wedge dx^m$ be an $m$-form of order $r$ on a jet space  $J^k(E)$ with $k \geq 2r$   and $\b$  the unique    source form   in $[d\a]$. Then  $\s:\cU\to E$
satisfies the variational principle of $\cI_{[\a]}$ if and only if, for any  $u\in \s^{(k)}(\cU) $ and $v \in T_uJ^k(E)|_\cU$,
\beq \label{Euler-Lagrange} \imath_{v}\b\big|_u(X_1,\ldots, X_m) = 0\qquad \text{for any choice of vectors}\ X_i \in T_u\big(\s^{(k)}(\cU)\big).\eeq
\end{theorem}
\begin{pf}
By Proposition \ref{TeoGusteaux} and the remark after \eqref{infinitesimalstationarity1},  a section $\s: \cU \to E$ satisfies the variational principle if and only if
\beq \label{acca} \int_{\s^{(k)}(D)}\imath_W \b = 0\eeq
for any regular domain $D \subset \cU$ and any $k$-th order variational field $W$. If  $D$ is sufficiently  small, so that $\s^{(k)}(D)$ is  included in the domain of a set of  adapted coordinates $\xi^{(k)} = (x^i, y^j, y^j_{I})$, we may  write
\beq W = W^j \frac{\partial}{\partial y^j} + \sum_{|I| = 1}^k W^j_{I} \frac{\partial}{\partial y^j_{I}} \quad \text{and}\quad \label {el}  \imath_W \b = (W^i \b_i) dx^1\wedge \ldots \wedge dx^m\ .\eeq
We also observe that,  for any given choice of maps $f^i: \s^{(k)}(D) \to \bR$, $ i = 1, \ldots, n$,  that vanish identically on a neighbourhood of $\p D$, one can construct   a smooth variation $F$  with  fixed boundary up to order $k$,  whose associated variational field $W$  has coordinate components  given by 
$$W^i|_{\s^{(k)}(p)}  = f^i|_{\s^{(k)}(p)}\qquad \text{for any} \ p \in D\ .$$
This  and \eqref{el} yield that \eqref{acca} is satisfied for any regular domain $D$ and any choice of   $W$ if and only if the restrictions $\b_i|_{\s^{(k)}(\cU)}$ are identically vanishing.  This means that $\s$ is a solution if and only if the $m$-form 
$$ \imath_W \b= (-1)^m W^i \b_i dx^1\wedge \ldots \wedge dx^m$$
 is identically vanishing on $\s^{(k)}(\cU)$ for {\it any} choice of a vector field $W = W^j \frac{\partial}{\partial y^j} + \sum_{|I| = 1}^k W^j_{I} \frac{\partial}{\partial y^j_{I}}$ at the points of $\s^{(k)}(\cU)$ (and  {\it not just for vector fields $W$ of variational type}). The claim follows.
\end{pf}
From this proof and the remarks before \eqref{2.12},  we directly see  that whenever the action $\cI_{[\a]}$ is determined by a Lagrangian density $L$, condition \eqref{Euler-Lagrange} holds if and only if the section $\s$ satisfies the  usual system of Euler-Lagrange equations determined by  $L$, as claimed.\par
\bigskip
\section{A new proof  of the Noether Theorem}
\label{NoetherTheoremsSection}
\setcounter{equation}{0}
\subsection{Conservation laws  for a system of  variational p.d.e.'s}\hfill\par
\label{subsect41}
 We call  {\it  $p$-form-valued differential operator of order $k$} a smooth  map
$\eta:J^k(E)\lra \L^{p}T^*\bR^m$  which makes  the following diagram commute
(here,  $\pi:\L^pT^*\bR^m \lra \bR^m $ is the standard projection) \par
 \centerline{\begin{picture}(110, 90)(0,0)
\put(0,70){$J^k(E)$}
\put(13,65){\vector(1,-1){35}}
\put(47, 15){$\bR^m$}
\put(53, 80){$\h$}
\put(33, 73){\vector(1,0){40}}
\put(80, 70){$\L^p T^* \bR^m$}
\put(95, 65){\vector(-1,-1){35}}
\put(85, 43){$\pi$}
\put(0, 43){\small $\pi_{k,-1}$}
\end{picture}}
Any such map has necessarily  the form \eqref{1.8} for some appropriate  $m$-tuple $P = (P^j)$ of smooth maps $P^j:  J^k(E) \to \bR$.  We call $P$ the 
{\it $m$-tuple of differential operators associated with $\h$}.  Note also that,  for any section $\s:\cU  \lra E$ defined on an open set $\cU \subset \bR^m$, 
the  map
$ \h|_{\s^{(k)}}\=\h\circ \s^{(k)}: \cU \lra \L^p T^* \bR^m$
is  a smooth $p$-form on $\cU \subset \bR^m$. 
\par
\medskip
Consider an $(m-1)$-valued differential  operator  $\h:J^k(E)\lra \L^{m-1}T^*\bR^n$   and  a variational class $[\a]$ of $m$-forms on $J^k(E)$.  We say that $\h$  satisfies  {\it a conservation law  for  $\cI_{[\a]}$}  if   for any section  $\s:\cU\lra E$ that   satisfies     the variational principle of  $\cI_{[\a]}$,    one has
\beq \label{def31}\int_{\p D} \h|_{\s^{(k)}}=0\eeq
on all  boundaries $\p D$ of regular regions  $D$ in the domain $\cU$ of $\s$.  One can directly check that this holds if and only if the associated $m$-tuple $P = (P^j)$ satisfies \eqref{1.2} 
for all solutions of the variational principle. 
\par
\smallskip
We now want to express  condition \eqref{def31} in terms of variational classes. Let    $\wt \h$ be the $(m-1)$-form on $J^k(E)$ defined by
\beq \label{def32} \wt \eta\big|_u \= (\pi^k_{-1})^*\eta(u) \qquad \text{for any}Ê\ u  \in J^k(E)\ .\eeq
By construction,  for any section  $\s:\cU\lra E$ and any  regular domain  $D\subset \cU$,
$$\int_{\p D} \h|_{\s^{(k)}} =\int_{\s^{(k)}(\p D)}\wt \eta\ .$$
Further, for any $(m-1)$-form $\wt \h'$, which is  in the same variational class of $\wt \h$ (i.e.  $\wt \eta' = \wt \eta + \l+d\mu$ for some  $\l$, $\mu$ holonomic),  we have
$$\int_{\s^{(k)}(\p D)}\wt \eta' =\int_{\s^{(k)}(\p D)}\wt \eta +\int_{\s^{(k)}(\p D)}\l+\int_{\s^{(k)}(D)}d^2\mu\overset{\l\ \text{is holon.}}=\int_{\s^{(k)}(\p D)}\wt \eta.$$
This shows that  \eqref{def31}  can be actually identified with an integral that   depends only of  the variational class  of   \eqref{def32}.  Conversely,  given an  arbitrary   $(m-1)$-form $\wt \h'$ on $J^k(E)$ and an  open set $\cU \subset \bR^m$  for which  one can determine  adapted coordinates on $\cW \= J^k(E|_{\cU})$,  one can  directly determine an $(m-1)$-form-valued differential operator $\h: J^k(E|_{\cU})\lra \L^{m-1}T^*\bR^n$ such that the  $(m-1)$-form   \eqref{def32} and  the restriction   $\wt \h'|_{(\pi^k_{-1})^{-1}(\cU)}$ are in the same variational class. These observations   motivate the following
\begin{definition} Let $\a$ and $\h$ be an $m$-form and an $(m-1)$-form, respectively, on $J^k(E)$. We say that the variational class $[\h]$
satisfies a {\it conservation law for  the action $\cI_{[\a]}$} if  for any section  $\s:\cU\lra E$ that satisfies   the variational principle of  $\cI_{[\a]}$,   one has  $\int_{\s^{(k)}(\p D)}\eta = 0$ for any  regular domain  $D\subset \cU$.
\end{definition}
By previous remarks, the  conservation laws satisfied by   $(m-1)$-form-valued differential operators determine  conservation laws satisfied by variational classes of $(m-1)$-forms. At a local level,  the converse is also true. \par
\bigskip
\subsection{\texorpdfstring{$\cI$}\ -Symmetries}Ê\hfill\par 
As announced in \S \ref{sect224}, our version of Noether Theorem is based on the following notions of ``symmetry''. \par
\begin{definition}\label{defDsymmetries} Let  $X$  and $\a$ be a vector field and an  $m$-form, respectively, defined  on an open subset $\cU$ of  $J^k(E)$.
\begin{itemize}
\item[a)] $X$ is an {\it infinitesimal  symmetry  of  the holonomic distribution  $\cD$} (shortly,  {\it $\cD$-symmetry}) if for all  holonomic vector field $Y$ on $\cU$,  the  Lie derivative $\cL_XY$ is  a holonomic vector field. 
\item[b)] $X$ is a {\it  weak $\cD$-symmetry} if,  for any  holonomic vector field $Y$ on $\cU$ and any $ u \in \cU$,  there exists a neighbourhood $\cU' \subset \cU$ of $u$ and a holonomic vector field $Y'$ on $\cU'$ such that 1)  $\pi^k_{k-1}{}_*(Y') = 0$ and 2) the Lie derivative  $\cL_X(Y- Y')$ is  holonomic.  
\item[c)] $X$ is an  {\it infinitesimal  (weak) symmetry  for   $\cI_{[\a]}$}  (shortly, {\it (weak) $\cI$-symmetry}) if
 it  is   a  (weak) $\cD$-symmetry
and  $\cL_X \a$ is holonomic for some proper $\a \in [\a]$. 
\end{itemize}
\end{definition}
The notion of   $\cD$-symmetry is the direct generalisation of the corresponding definition  considered in \cite{FS}.  There,  the discussion was limited  to the case of jet spaces of maps of  one independent variable, but most of their properties remain true in our more general situation.   We briefly  recall the main properties of $\cD$-symmetries  and refer to \cite{FS} for  other details. \par
\begin{itemize}
\item[1)] If a vector field $X$ on an open subset  $\cU \subset J^k(E)$ is a $\cD$-symmetry, then its local flow  is a 1-parameter family  of local diffeomorphisms  mapping any holonomic submanifold  $\s^{(k)}(\cU)$  into another   submanifold, which is also locally holonomic, i.e. of the form $\s'{}^{(k)}(\cU')$ for some other section $\s': \cU' \to E$. \par
\item[2)] If  $X$ is a $\cD$-symmetry and $\l$ is  a holonomic $p$-form, then the local flow $\Phi^X_t$ of $X$ is such that all  local  $p$-forms $\Phi^X_t{}^*(\l)$, $t \in (-\ve, \ve) \subset \bR$, are holonomic. Hence,   also the  Lie derivative $\cL_X \l$ is holonomic.
\item[3)]ÊIf $\a$, $\a'$ are in the same variational class  (i.e. $\a - \a' = \l + d \mu$, with $\l$, $\mu$ holonomic), then
$\cL_X \a$ is holonomic if and only if $\cL_X \a'$ is holonomic.
 \end{itemize}
 The class of weak $\cD$-symmetries is new and it naturally includes all $\cD$-symmetries.  This weaker version  of the $\cD$-symmetries  is needed to remove  an  incorrect claim of  \cite{FS} (see  Appendix). We remark that  if  one  works on the infinite order  jet space $J^\infty(E)$   instead of   the finite order jet space $J^k(E)$,   the   notions of $\cD$-symmetry and weak $\cD$-symmetry   coincide.\par
\medskip
\subsection{Differential equations that characterise  \texorpdfstring{$\cD$}\ -symmetries and infinitesimal symmetries of an action}\hfill \par
\begin{prop}\label{equiv-Dsimm}
Let $X$ and $\a$ be a vector field and an  $m$-form, respectively,  on  an open subset $\cU' \subset J^k(E)$ and  assume that $\wh \xi^{(k)} = (x^i, y^j, y^j_{J})$  are adapted coordinates on $\cU  \subset \cU'$. Then:
\begin{itemize}
\item[(1)] A necessary condition for $X|_{\cU}$  to be a  $\cD$-symmetry  is that it satisfies the differential equations
\beq
 \label{Dsymmetrieseq}
 \o^i_{I}\left(\cL_X \frac{d}{dx^j}\right) = 0
 \eeq
 for all
 $1 \leq j \leq m$, $1 \leq i  \leq n$ and  $0 \leq |I| \leq k-1$. Conversely, if $X$ satisfies  the above system of differential equations, then it  is a weak $\cD$-symmetry.
 \item[(2)] A necessary condition for $X|_{\cU}$ to be  an  infinitesimal symmetry for $\cI_{[\a]}$ (considered as  functional  on  the sections of $E|_{\pi^k_{0}(\cU)}$) is that for some $\a_o \in [\a|_{\cU}]$  it satisfies
the system of differential equations \eqref{Dsymmetrieseq} together with  the differential equations
\beq
  \begin{split} \label{symmetries}
&(\cL_X \a_o)\left( \frac{d}{d x^{i_1}},\ldots, \frac{d}{d x^{i_{m-r}}}, \frac{\partial}{\partial y^{j_1}_{J_1}}, \ldots,  \frac{\partial}{\partial y^{j_r}_{J_r}}\right)   = 0
\end{split}
\eeq
for all   $0 \leq r \leq m-1$, $1 \leq i_h \leq m$, $1 \leq j_\ell\leq n$ and $|J_\ell| = k$.  Conversely, if $X$ satisfies the systems \eqref{Dsymmetrieseq} and \eqref{symmetries} for some $\a_o \in [\a|_\cU]$, then $X|_{\cU}$ is a infinitesimal weak symmetry for $\cI_{[\a]}$.
 \end{itemize}
\end{prop}
\begin{pf} (1) Recall that $\cD|_{\cU}$ is generated by the vector fields $\frac{d}{dx^i}$, $1 \leq i \leq m$,  and the vector fields $\frac{\partial}{\partial y^j_{J}}$,  $1 \leq j \leq n$, $|J| = k$. It therefore consists of the intersections of the kernels  of  the  1-forms $\o^i_I$, $0 \leq |I| \leq k-1$, $1 \leq i \leq n$, at all points $u \in \cU$. Hence $X|_{\cD}$ is a $\cD$-symmetry  only if  \eqref{Dsymmetrieseq} holds. Conversely, assume that  $X$ satisfies the system of differential equations \eqref{Dsymmetrieseq} and let  $Y$ be a  holonomic vector field, i.e.  
$$Y = Y^j \frac{d}{dx^j} + \sum_{1 \leq j \leq n, |J| = k} Y^i_J \frac{\p}{\p y^i_J}\ .$$
Then,   $Y' \= \sum_{1 \leq j \leq n, |J| = k} Y^i_J \frac{\p}{\p y^i_J}$ is a holonomic vector field  such that: a) $\pi^k_{k-1*}(Y') = 0$;  
b) $ \o^i_{I}\left(\cL_X (Y - Y')\right) = 0$ for any $1 \leq i \leq m$ and  $1 \leq |I| \leq k-1$. This means that   $\cL_X (Y - Y')$ is holonomic and proves that $X$ is a weak $\cD$-symmetry.\par
(2) From (1) and definition of  infinitesimal symmetry for $\cI_{[\a]}$, it follows that the systems  \eqref{Dsymmetrieseq} and \eqref{symmetries}  are necessary conditions for $X$ to be a $\cD$-symmetry. The  converse claim is a  consequence of  definitions and  (1).
\end{pf}
\subsection{Explicit expressions for weak \texorpdfstring{$\cD$}\ -symmetries}\hfill\par
Let  $\wh\xi^{(k)} = (x^j, y^i,y^i_{J})$ be a       system of adapted coordinates
 on an open subset  $\cU \subset J^k(E)$. For a given  smooth map
$$\vv = (\vvb^1,  \ldots, \vvb^m,  \vv^1, \ldots, \vv^n): \cU \subset J^k(E) \longrightarrow \bR^{m+n}\ ,$$
let us   denote by    $X_\vv$   the   vector field
\beq \label{Xv}
X_\vv \= \vvb^j\frac{\partial}{\partial x^j} + \vv^i\frac{\partial}{\partial y^i}
+ \sum_{1\leq i \leq n, 1 \leq |J|\leq k} \vv^i_{J}\frac{\partial}{\partial y^i_{J}}\ ,
\eeq
where we follow the standard Einstein convention on summation over repeated indices and, for any multiindex $J = (J_1, \ldots, J_m)$ with $1 \leq |J| \leq k$, we set 
\beq \label{Xvbis}
\vv^i_{J} \=  \left( \frac{d}{d x^1}\right)^{J_1}\!\!\!\!{\ldots} \left(\frac{d}{d x^m}\right)^{J_m}\left(\vv^i-y^i_{1_r} \vvb^r\right)+y^i_{J + 1_r}\vvb^r\, .
\eeq
(in this formula, we  assume $y^i_{J} \= 0$ when $|J| = k+1$).
This  yields that 
 \begin{multline} \label{Xvter}
X_\vv = \vvb^j \frac{d}{d x^j} + \left(\vv^i- y^i_{1_r} \vvb^r\right)\frac{\partial}{\partial y^i}
+\\
+ \sum_{1 \leq |J| \leq k} \left( \frac{d}{d x^1}\right)^{J_1}\!\!\!\!{\ldots} \left(\frac{d}{d x^m}\right)^{J_m}\!\!\!\left(\vv^i - y^i_{1_r} \vvb^r\right)\,\frac{\partial}{\partial y^i_{J}}\ .
\end{multline}
\medskip
We may now prove the following
\begin{prop} \label{lemma61}  Given a set of adapted coordinates  $\wh\xi^{(k)} = (x^j, y^i,y^i_{J})$
 on an open subset  $\cU \subset J^k(E)$, 
 the  vector fields  $X_\vv$ defined in \eqref{Xv}  are exactly the vector fields that satisfy the system of differential equations  \eqref{Dsymmetrieseq}. In particular, all of them   are  weak $\cD$-symmetries.
\end{prop}
\begin{pf} Let $\displaystyle
X = \Xb^j\frac{\partial}{\partial x^j} + X^i\frac{\partial}{\partial y^i}
+ \sum_{1 \leq i \leq n, 1 \leq |J| \leq k} X^i_{J}\frac{\partial}{\partial y^i_{J}}
$. 
   Since
$$
\cL_X \frac{d}{dx^j} =  - \frac{d \Xb^r}{dx^j}\frac{\partial}{\partial x^r} + \sum_{0 \leq |J| \leq k -1}\left(X^i_{J+ 1_j}
- \frac{d X^i_{J}}{dx^j}\right)\frac{\partial}{\partial y^i_{J}} - \sum_{|J| = k}\frac{d X^i_{J}}{dx^j} \frac{\partial}{\partial y^i_{J}}\ ,
$$
the system  \eqref{Dsymmetrieseq} is equivalent to 
\beq \label{Omega=0}
0=\omega^i_{I}\left(\cL_X \frac{d}{dx^j} \right) = X^i_{I + 1_j} - \frac{d X^i_{I}}{d x^j} + y^i_{I + 1_r} \frac{d \Xb^r}{dx^j}
\eeq
for all  $ 1 \leq i \leq n$ and $0 \leq |I| \leq k-1$.  This means that  if $X$ is  a solution  to  \eqref{Dsymmetrieseq}, then the  components $X^i_{J}$ with $1 \leq |J| \leq k$  are uniquely  determined by an inductive process from  the components $\Xb^r$ and $X^j$. If we set $\vvb^r \= \Xb^r$ and $\vv^j \= X^j$, 
a straightforward check shows that $X = X_\vv$.  The last claim follows from Proposition \ref{equiv-Dsimm}.
\end{pf}
\par
\bigskip
\subsection{The Noether Theorem}\hfill\par
\label{section45}
\begin{definition}\label{formaradicale}
We say that an $m$-form $\a$ on an open set $\cU \subset J^k(E)$   is  {\it of Poincar\'e-Cartan type} if its differential $d \a$ is equal to a source form up to addition of an holonomic $(m+1)$-form.
\end{definition}
The main motivation for this  terminology comes from the fact that  the well-known   Poincar\'e-Cartan form $\a = p_i dq^i - H dt$ of Hamiltonian Mechanics is a $1$-form of Poincar\'e-Cartan type according to the above definition (see \cite{FS} for details;  see also  \cite{An}, Ch. 5B,  for other generalisations of the Poincar\'e-Cartan 1-form). 
We also  remark  that  if  $\a_L$ is an $m$-form on $\cU \subset J^k(E)$ with adapted coordinate expression $\a_L = L dx^1 \wedge \ldots \wedge d x^m$  for some Lagrangian $L$ of order $r \leq \left[\frac{k}{2}\right]$,   then for  any $u \in \cU$  there exists a neighbourhood $\cU' \subset \cU$ of $u$ such that  the variational class  $[\a_L|_{\cU'}]$  contains at last one  1-form of Poincar\'e-Cartan type. To see this,
 one needs only to consider   a system of adapted coordinates on a  neighbourhood $\cU'$ of $u$ and   the source form $\b \in  [d\a|_{\cU'}]$  in \eqref{sorgente}, which has  components determined by the Euler-Lagrange operator applied to $L$.  Then $\b = d \a_L|_{\cU'}+ d \mu + \l = d(\a_L|_{\cU'}+ \mu) + \l$  for some holonomic  $\mu$ and $\l$  and $\a \= \a_L|_{\cU'} + \mu$
is the required  $m$-form of Poincar\'e-Cartan type in $[\a_L|_{\cU'}]$.\par
\smallskip
We finally observe that  the previous argument shows that if $L$ is a Lagrangian of order $r$ and $k > 2r$,   the variational class $[\a_L|_{\cU'}]$ contains an $m$-form which is not only  of Poincar\'e-Cartan type, but also  {\it of order $k_o\leq k-1$}.  This additional condition  is quite useful and it will  be often required in the following.
\par
\medskip
The notion of $m$-forms of Poincar\'e-Cartan type leads to the following  characterisation of weak $\cI$-symmetries.
As in Proposition \ref{lemma61},   we consider a fixed adapted coordinates
$\wh\xi^{(k)} = (x^j, y^i,y^i_{(a)})$ on an open set  $\cU \subset J^k(E)$.
\begin{prop} \label{lemma62} Let $\a$ be an m-form of Poincar\'e-Cartan type  of order $k_o \leq k-1$ and $
X$ a weak $\cD$-symmetry on $\cU \subset J^k(E)$ satisfying the system \eqref{Dsymmetrieseq}. 
Then $X$ is such that $\cL_X \a$ is holonomic (thus, an  infinitesimal weak symmetry for $\cI_{[\a]}$)  if and only if  it  satisfies the following system of linear differential equation  for some source form  $\b \in [d \a|_{\cU}]$:
\begin{align}  
\nonumber &  \cL_{X} \a\left(\frac{d}{dx^{i_1}}, \ldots, \frac{d}{dx^{i_r}}, \frac{\partial}{\partial y^{j_1}_{J_1}}, \ldots, \frac{\partial}{\partial y^{j_{m-r}}_{J_{m-r}}}\right) = \\
\label{3.25ter} & \qquad \qquad \qquad = d(\imath_{X} \a)\left(\frac{d}{dx^{i_1}}, \ldots, \frac{d}{dx^{i_r}}, \frac{\partial}{\partial y^{j_1}_{J_1}}, \ldots, \frac{\partial}{\partial y^{j_{m-r}}_{J_{m-r}}}\right) =  0, 	\\	
  \label{3.25bis}& d(\imath_{X} \a)\left(\frac{d}{d x^1},\ldots, \frac{d}{d x^m} \right)  = -  \b\left(X, \frac{d}{d x^1},\ldots, \frac{d}{d x^m} \right)
  \end{align}
for all  $1 \leq r \leq m-1$  and all choices of indices and multiindices with $1 \leq i_h \leq m$, $1 \leq j_\ell\leq r$ and $|J_\ell| = k$.
\end{prop}
\begin{pf} Let  $\l$ be the   holonomic $(m+1)$-form   $\l \= d \a - \b $.
From definitions,  for any $0 \leq r \leq m$, $1 \leq i_\ell \leq m$,  $1 \leq j_h \leq n$, $|J_h| = k$,  we have
$$ \l \left(\frac{d}{dx^{i_1}}, \ldots, \frac{d}{dx^{i_r}}, \frac{\partial}{\partial y^{j_1}_{J_1}}, \ldots, \frac{\partial}{\partial y^{j_{m-r}}_{J_{m-r}}},\star\right) = 0\ ,\qquad
\imath_{\frac{\partial}{\partial y^{j_1}_{J_1}}} \b = 0\ .$$
 From this and Proposition  \ref{equiv-Dsimm} (2), it follows  that  the weak $\cD$-symmetry  $X$  is an infinitesimal symmetry for $\cI_{[\a]}$ if and only if  
 \eqref{3.25ter} and  \eqref{3.25bis} hold. 
\end{pf}
We can now state and prove the Noether Theorem in its two parts, direct and inverse.
\begin{theo}[\bf Noether Theorem -- first part]\label{directNoether}
Let $\a$ be an $m$-form of Poincar\'e-Cartan type on $J^k(E)$.
 If  $X$  is   a (weak) infinitesimal symmetry for $\cI_{[\a]}$ on an open set $\cU \subset J^k(E)$ with $\cL_X \a$ holonomic, then the  variational class of the $(m-1)$-form
 $\h^{(X)} \= \imath_X \a$
  satisfies a conservation law  for the  action  $\cI_{[\a]}$.
\end{theo}
\begin{pf} Since $\a$ is of Poincar\'e-Cartan type, there is a holonomic $(m+1)$-form $\l$ such that   $\b = d \a + \l$ is a source form.
Hence, if  $\cL_X \a$ is holonomic,  then for any section  $\s: \cV \subset \bR^m \longrightarrow \cU \subset E$ and any regular domain $D \subset \cV$
\begin{multline} \int_{\s^{(k)}(\p D)}\eta^{(X)} =     \int_{\s^{(k)}(D)} d( \imath_X \a) = \\
 =
\int_{\s^{(k)}(D)} \cL_X \a -  \int_{\s^{(k)}(D)} \imath_X d \a \ \ \overset{\cL_X \a\ \text{and}\ \l\ \text{holon.}}=  \  -  \int_{\s^{(k)}(D)} \imath_X\b \ .
\end{multline}
Since    
$ \int_{\s^{(k)}(D)} \imath_X\b = 0$ for any solution $\s$  of the variational principle,   $[\h^{(X)}]$ satisfies a conservation law.
\end{pf}
Now, before   getting into the  second part of Noether Theorem, we need to introduce an appropriate definition  of regularity for
Euler-Lagrange equations.\par
\smallskip
Let $[\alpha]$ be a variational class of $m$-forms  on an open subset of  $J^k(E)$
and assume that $\b = \b_i \o^i_{0} \wedge dx^1\wedge \ldots \wedge dx^m$  is a source form  on some open subset $\cW \subset J^k(E)$. Assume also that $\b$ is
 of order $k_\b \leq k-1$ (we may always reduce to this case by pulling back $\b$ on some jet space of higher order) and   consider  the differentials $d\b_i$  of the components $\b_i$ of $\b$. By  assumptions,
 these differentials are equal to
$$d \b_i = \sum_{j=1}^m\frac{\partial \b_i}{\partial x^j} dx^j +  \sum_{\begin{subarray}{c} 1 \leq j \leq n\\0 \leq |I| \leq k-1\end{subarray}}\frac{\partial \b_i}{\partial y^j_{I}} dy^j_{I} =  \sum_{j=1}^m\frac{d \b_i}{dx^j} d x^j +  \sum_{\begin{subarray}{c} 1 \leq j \leq n\\0 \leq |I| \leq k-1\end{subarray}}\frac{\partial \b_i}{\partial y^j_{I}} \o^j_{I}\ .$$
Due to this, for any  section $\s: \cU \to E$ whose $k$-th order lift $\s^{(k)}$ takes values in $\cW$,  we have
$$d\left(\b_i(\s^{(k)}(x^1, \ldots, x^m))\right) =  d \b_i\left( \s^{(k)}_*\bigg(\frac{\p}{\p x^j}\bigg)\right) = \sum_{j=1}^m \left.\frac{d \b_i}{dx^j} \right|_{\s^{(k)}(t)}d x^j\ .$$
Hence,  $\s$ is  a solution of the Euler-Lagrange  equations, i.e. 
\beq \label{firstsystem} \b_i(\s^{(k)}(x^1, \ldots, x^m)) = 0\ ,\qquad 1 \leq i \leq n\ ,\eeq
 if and only if it is also  a solution to  the (expanded) system 
\beq \label{firstprol} \b_i(\s^{(k)}(x^1, \ldots, x^m)) =  \frac{d \b_i}{dx^j}(\s^{(k)}(x^1, \ldots, x^m))= 0\ ,\ 1 \leq j \leq m, 1 \leq i \leq n.\eeq
The system   \eqref{firstprol} is  called {\it first  prolongation} of  \eqref{firstsystem}. 
Note that   {\it if the $0$-forms (= functions)  $\b_i$  are  of order $k_\b$ ($\leq k - 1$),  then, generically, the   functions   that define  \eqref{firstprol} are  $0$-forms of order  $k_\b + 1$}. \par
Iterating this argument   $(k'-k_\b)$ times for some $k' \leq k$,  we get  that  \eqref{firstsystem} is equivalent to the expanded system
\begin{multline} \label{pthprol} \b_i(\s^{(k)}(x^\ell)) =  \frac{d \b_i}{dx^{j_1}}(\s^{(k)}(x^\ell)) = \ldots  =\\
=  \left(\frac{d}{dx^{j_1}}\bigg(\frac{d}{dx^{j_2}}\ldots \bigg(\frac{d}{dx^{j_{k'  - k_\b}}}(\b_i)\bigg)\ldots\bigg)\right)(\s^{(k)}(x^\ell))  =0\end{multline}
for all $1 \leq j_h \leq m$ and $1 \leq i \leq n$.
 This new system is called    {\it  full  prolongation of
\eqref{firstsystem} up to order $k' $}.  Note that,   generically, the   $0$-forms  that give  the   full prolongation up to order $k' $  are {\it $0$-forms of order  $k' $}.
\par
\begin{definition} For a given  system of Euler-Lagrange equations \eqref{firstsystem} of order $k_\b$, let  $F^{(k' )}_\b$ 
be the smooth map
\beq \label{effesse} F^{(k' )}_\b: \cW \subset J^k(E)\to \bR^N\ ,\qquad F^{(k' )}_\b \= \bigg( \b_i,  \frac{d^{|I|}\b_i}{dx^I} \bigg)_{\ \begin{subarray}{l} 1 \leq i \leq m\\ 1 \leq  |I| \leq k'  - k_\b
\end{subarray}}\!\!\!\!.\eeq
Here, $N = n\left(1 + \sum_{r = 1}^{k' - k_\b} \left(\smallmatrix m+r-1\\ r\endsmallmatrix\right)\right)$ and $\frac{d^{|I|} \b_i}{dx^I} $, $I = (I_1, \ldots, I_m)$,  stands for  
$$\frac{d^{|I|} \b_i}{dx^I} \=\underset{I_1\text{-times}}{\underbrace{ \frac{d}{dx^1}  \frac{d}{dx^1}  \dots  \frac{d}{dx^1} }}\,\underset{I_2\text{-times}}{\underbrace{ \frac{d}{dx^2}  \frac{d}{dx^2}  \dots  \frac{d}{dx^2} }}\dots \underset{I_m\text{-times}}{\underbrace{ \frac{d}{dx^m} \frac{d}{dx^m} \ldots  \frac{d}{dx^m}} }\b_i\ .$$
Let  $Z^{(k')}_\b  \= \{\ u \in \cW\ : \ F^{(k')}_\b(u) = 0\ \} \subset J^k(E)$. We say that the system of Euler-Lagrange equations \eqref{firstsystem} is  {\it   $k'$-regular on $\cW$} if:
\begin{itemize}
\item[i)] the $k'$-th order jets of the solutions of the Euler-Lagrange equations constitute a dense subset of $\pi^k_{k'}(Z^{(k')}_\b) \subset J^{k'}(E)$ and   
\item[ii)] the map  $F^{(k')}_\b$ is a submersion at all points of $Z^{(k')}_\b$. 
\end{itemize}
\end{definition}
We are now able to state and prove the second part of Noether Theorem.
\begin{theo}[\bf Noether Theorem -- second part] \label{inverseNoether} Let $k_o \leq \left[\frac{k}{2}\right]-1$ and $\a$ be an  $m$-form of Poincar\'e-Cartan type on  $J^k(E)$ of order less than or equal to  $k_o-1$. Assume that  there exists an open subset $\cW \subset J^k(E)$ admitting   a  system of adapted coordinates  $\wh\xi^{(k)} = (x^j, y^i,y^i_{I})$, where the  following non-degeneracy conditions are satisfied:
\begin{itemize}
\item[a)]    the  source form in $ [d \a|_{\cW}]$ 
$$\b = \b_i dy^i  \wedge dx^1 \wedge \ldots \wedge dx^m = \b_i \o^i_{0} \wedge dx^1 \wedge \ldots \wedge dx^m$$
 is of order $k_\b \leq k_o$  and    the system of  Euler-Lagrange equations  \eqref{firstsystem} is $k_o$-regular on $\cW$;
\item[b)] $\left.\a\left(\frac{d}{dx^1}, \ldots, \frac{d}{dx^m}\right)\right|_{u}  \neq 0$ at all $u$'s in $\cW$.
\end{itemize}
 \par
Then, if $\h$ is an $(m-1)$-form on $\cW$ of order less than or equal to  $k_o-1$, such that $[\h]$ satisfies a conservation law for    $\cI_{[\a]}$,   there exists a neighbourhood $\cU$ of $Z^{(k_o)}_\b  = \{u \in \cW\ : \ F^{(k_o)}_\b(u) = 0\}$ on which there are 
\begin{itemize}
\item[1)] a weak $\cI$-symmetry  $X$ for $\cI_{[\a]}$ with $\cL_X \a$ holonomic  and
\item[2)] an $(m-1)$-form   $\gz$ that vanishes   identically   on any  $(m-1)$-tuple of  vectors  in a tangent space of a holonomic submanifold  $\s^{(k)}(\cV)$ of   a solution  $\s$ to the variational principle,
\end{itemize}
such  that
\beq \label{noether2}   \h\big|_{\cU}  =  \imath_{X} \a + \gz\  .\eeq
\end{theo}
\begin{pf} By Propositions \ref{lemma61} and  \ref{lemma62},  it suffices to prove the existence of a smooth $\bR^{m+n}$-valued map $\vv = (\vvb^j, \vv^i): \cU \to \bR^{m+n}$ on a neighbourhood $\cU$ of $Z^{(k_o)}_\b$ and  of  an $(m-1)$-form $\gz$   on $\cU$, such that: i) $\gz(Y_1, \ldots, Y_m) \big|_{\s^{(k)}(\cV)} = 0$ for any choice of vector fields $Y_i$ tangent  to submanifolds    $\s^{(k)}(\cV)$ of  solutions $\s$ of the Euler-Lagrange equations; ii)    the following equations are  satisfied
\begin{align}\label{ohoh}
 &\imath_{X_\vv}\a  = \h - \gz \ ,\\
 &\b\left(X_\vv, \frac{d}{d x^1},\ldots, \frac{d}{d x^m} \right) {=} -
\label{ohoh1} d  \h\left(\frac{d}{d x^1},\ldots, \frac{d}{d x^m} \right) + d \gz \left(\frac{d}{d x^1},\ldots, \frac{d}{d x^m}\right)\ ,\\
\nonumber &  \cL_{X_\vv} \a \left(\frac{d}{dx^{j_1}}, \ldots, \frac{d}{dx^{j_r}}, \frac{\partial}{\partial y^{j_1}_{J_1}}, \ldots, \frac{\partial}{\partial y^{j_{m-r}}_{J_{m-r}}}\right) = \\
\label{ohoh2}  & \qquad  = d(\imath_{X_\vv} \a)\left(\frac{d}{dx^{j_1}}, \ldots, \frac{d}{dx^{j_r}}, \frac{\partial}{\partial y^{j_1}_{J_1}}, \ldots, \frac{\partial}{\partial y^{j_{m-r}}_{J_{m-r}}}\right) =  0
 \end{align}
for all $1 \leq r \leq m-1$,    $1 \leq i_h \leq m$, $1 \leq j_\ell\leq r$ and $|J_\ell| = k$.
Let us write $\a$, $\h$, $\gz$ and $\b$  as sums of homogeneous forms, that is as 
\begin{align}
\label{lillina} \a &= \a_0 dx^1\wedge \ldots \wedge dx^m  +\\
\nonumber &\qquad \qquad \qquad  + \sum_{\smallmatrix 0 \leq |I| \leq k-1\\ 1 \leq i \leq n, 1 \leq j \leq m\endsmallmatrix} \a^{I}_{i|j} \o^i_{I}\wedge dx^1 \wedge \ldots  \underset{j}{\wh{\phantom{dx^i}}} \ldots \wedge dx^m + \l^{(\a)}
\end{align}
\begin{align}
\label{lillina1} \h &= \sum_{1 \leq j \leq m}\h_j dx^1 \wedge \ldots  \underset{j}{\wh{\phantom{dx^i}}}\ldots \wedge dx^m  + \mu^{(\h)}\\
\label{lillina2} \gz &=\!\!\!\sum_{1 \leq j \leq m}\!\!\!\gz_j dx^1 \wedge \ldots  \underset{j}{\wh{\phantom{dx^i}}}\ldots \wedge dx^m  + \mu^{(\gz)}\\
\label{lillina2bis}Ê\b & = \sum_{1\leq i  \leq n}\b_i \o^i_{0} \wedge dx^1 \wedge \ldots \wedge dx^m \ ,
\end{align}
where $\l^{(\a)}, \mu^{(\h)}, \mu^{(\gz)}$ are holonomic forms, given by the terms in $\a$, $\h$ and $\gz$, respectively,  that are homogeneous of bidegree $(\ell, h)$ with $\ell \leq m-2$.
We now observe that,  for any vector field $X$, the $(m-1)$-form $\imath_X \l^{(\a)}$ is   holonomic.  Hence, the $(m-1)$-form
 $\gz$ has the prescribed properties  if and only if 
 $\gz' \= \gz + \imath_X \l^{(\a)} -  \mu^{(\h)}$  has those  properties.  This yields that the above  conditions are satisfied by $\gz$, $\a$ and $\h$    if and only if they are satisfied by  $\gz'$,  $\a' = \a - \l^{(\a)}$ and $\h' = \h - \mu^{(\h)}$. Due to this,   {\it we may safely assume  $\l^{(\a)} = 0$ and $ \mu^{(\h)} = 0$}.\par
\smallskip
Consider now the function
$$g : \cW \longrightarrow \bR\ ,\qquad g(u) \= d  \h\left(\frac{d}{d x^1},\ldots, \frac{d}{d x^m} \right)\bigg|_u\ .$$
We claim that
 $g$ vanishes identically on $Z^{(k_o)}_\b$. Indeed, by assumption (a), the $k_o$-th order jets of  solutions $\s$ to the Euler-Lagrange equations form a dense subset $\wt Z$ of
  $\pi^k_{k_o}(Z^{(k_o)}_\b)$. In particular,  for any $\wt u \in \wt Z$ and $1 \leq i \leq m$, we have that 
$$\left.\frac{d}{dx^i}\right|_{\wt u} -   \s^{(k_o)}_*\bigg(\left.\frac{\p}{\p x^i}\right|_p\bigg) \in \text{Span}\left\{\left.\frac{\partial}{\partial y^j_{J}}\right|_u\ ,\ |J| = k_o\right\} \subset T_{\wt u} J^{k_o}(E)\ ,$$
for some solution $\s$ with   $ \s^{(k_o)}(p) = \wt u$.
Since $[\h]$ satisfies a conservation law and $\h$  is   of order less than or equal to $k_o -1<  k_o$, we get that for any jet $u \in (\pi^k_{k_o})^{-1}(\wt Z) \subset Z^{(k_o)}_\b$
$$g(u) \= d  \h\left(\frac{d}{d x^1},\ldots, \frac{d}{d x^m} \right) \bigg|_u\!\! {=}\, d\h\left(\s^{(k)}_*\bigg(\frac{\p}{\p x^1}\bigg|_{p}\bigg),\ldots, \s^{(k)}_*\bigg(\frac{\p}{\p x^m}\bigg|_{p}\bigg)\right) = 0.$$
By continuity of $g$, we get  $g(u) = 0$ for any $u \in Z^{(k_o)}_\b$. \par
\smallskip
Since we are also assuming that  $F^{(k_o)}_\b: \cW \longrightarrow \bR^N$ is a submersion at any  $u \in Z^{(k_o)}_\b$ and that $\h$ is of order less than  or equal to $k_o-1$, by    standard properties of  submanifolds (see e.g., \cite{Mi}, Lemma 2.1 and \cite{Ol}, Prop. 2.10),
there exist  an open neighbourhood  $\cU \subset \cW$ of $Z^{(k_o)}_\b$ and   some  (non-uniquely determined) smooth functions $\wh \vv^j_{I}$ on $\cU$,  with $1 \leq j \leq n$,  $0 \leq |I| \leq k_o - k_\b$,  such that
\begin{equation}
\label{acci} g =  d  \h\left(\frac{d}{d x^1},\ldots, \frac{d}{d x^m} \right) =  \sum_{\smallmatrix 0 \leq |I| \leq k_o -k_\b\\ 1 \leq j \leq n\endsmallmatrix}\wh \vv^j_{I} \frac{d^{|I|} \b_j}{dx^I} \ .
\end{equation}
Further, since   the left-hand side of \eqref{acci} is a function of the coordinates $(x^i, y^j_I)$ with $|I| \leq k_o$,  there is no loss of generality if we assume that all functions $\wh \vv^j_{I}$ depend only on the coordinates $(x^i, y^j_I)$ with $|I| \leq k_o$.\par
\smallskip
Now, if $k_o > k_\b$, let $\gz^{(1)}$ be the $(m-1)$-form on $\cU$ defined by
\beq \gz^{(1)} \= \sum_{\smallmatrix    |I| = k_o - k_\b\\ 1 \leq j \leq n, 1 \leq i \leq m\endsmallmatrix} (-1)^{i-1} \wh \vv^j_{I} \frac{d^{(k_o - k_\b -1)} \b_j}{dx^{I - 1_i}}dx^1 \wedge \ldots \underset{i}{\wh{\phantom{dx^i}}}  \ldots \wedge dx^m \ .
\eeq
By construction, the integrals of the  $(m-1)$-form and of its exterior differential $d \gz^{(1)}$ along  the holonomic submanifolds $\s^{(k)}(\cV)$ of solutions $\s: \cV \to E$  of the variational principle are identically equal to $0$. Indeed,  this is a consequence of the fact that, modulo holonomic terms,  the  components of $\gz^{(1)}$  and of $d \gz^{(1)}$  are given by   linear combinations of the   components of the map  $F^{(k_o)}_\b $.
 Furthermore, we have 
$$\sum_{\begin{subarray}{l}  |I|  =  k_o  - k_\b\\ 1 \leq j \leq n\end{subarray}}\wh \vv^j_{I} \frac{d^{k_o - k} \b_j}{dx^I} = d \gz^{(1)} \left(\frac{d}{d x^1},\ldots, \frac{d}{d x^m} \right) - g^{(1)} \qquad \text{with}$$
$$ g^{(1)}\= \sum_{\begin{subarray}{l} 
|I| = k_o - k_\b\\
1 \leq j \leq n, 1 \leq i \leq m
\end{subarray}} \frac{d \wh \vv^j_{I}}{d x^i} \frac{d^{k_o -k_\b -1} \b_j}{dx^{I - 1_i}} + \sum_{1 \leq |J| \leq k_o - k_\b - 1}\wh \vv^j_{I} c^J_{i I-1_i}  \frac{d^{|J|} \b_j}{dx^{J}}\ ,$$
where we used the notation  $c^J_{i I-1_i} $ to indicate the  unique functions  that satisfy the identity
$$\frac{d^{|I|}}{dx^{I}} =  \frac{d}{dx^i} \frac{d^{|I|-1} }{dx^{I- 1_i}} +\sum_{0 \leq |J| \leq |I|-1} c^J_{i I-1_i}  \frac{d^{|J|}}{dx^{J}}\ .$$
Combining this with \eqref{acci}, we get  the existence of  functions $\wh \vv'{}^j_I$  on $\cU$, which   depend  only on the coordinates $(x^i, y^j_I)$ with $|I| \leq k_o+1$ and  such that 
\beq\label{accibis}    d  (\h- \gz^{(1)})\left(\frac{d}{d x^1},\ldots, \frac{d}{d x^m} \right) =\sum_{\smallmatrix 0 \leq |I| \leq k_o - k_\b -1\\ 1 \leq j \leq n\endsmallmatrix}\wh \vv'{}^j_{I} \frac{d^{|I|} \b_j}{dx^I} \ .
 \eeq
Iterating this construction, we obtain a sequence  of $(m-1)$-forms $\gz^{(1)}$, $\gz^{(2)}$, \ldots, $\gz^{(k_o - k_\b)}$ such that 
\beq\label{acciter}  d  (\h- \gz^{(1)} - \ldots - \gz^{(k_o - k_\b)})\left(\frac{d}{d x^1},\ldots, \frac{d}{d x^m} \right) =  \sum_{i = 1}^n \wh \vv^i \b_i \ ,\eeq
for some appropriate smooth functions $\wh \vv^i:  \cU \longrightarrow \bR$ on an open neighbourhood  $\cU$ of $Z^{(k_o)}_\b$, which depend only on the coordinates $(x^i, y^j_I)$ with $|I| \leq 2 k_o- k_\b \leq 2 k_o$. Note that, 
by  construction of  such $(m-1)$-forms $\gz^{(i)}$,    their   integrals along   holonomic submanifolds $\s^{(k)}(\cV)$ of solutions $\s: \cV \to E$  to the variational principle, are identically equal to $0$. \par
\medskip
Set $\gz =   \gz^{(1)} + \ldots + \gz^{(k_o -k_\b)}$ and,  for any $i = 1, \ldots, m$, denote by $h_i$ the function 
$$\ h_i\= \h_i - \gz_i - \sum_{\smallmatrix 0 \leq J\leq |k - k_o -1|\\ 1 \leq j \leq n\endsmallmatrix} \gz\bigg(\frac{\p}{\p y^j_I}, \frac{d}{d x^1}, \ldots \underset{i}{\wh{\phantom{a}}}\ldots \frac{d}{d x^m}\bigg)$$
where we denote by $\h_i$ and $\gz_i$  the coefficients of   the $(m-1)$-forms $dx^1 \wedge \ldots  \underset{i}{\wh{\phantom{dx^i}}}\ldots \wedge dx^m$ in the expansions of  $\h$ and $\gz$. 
Then, consider the $(m+n)$-tuple of functions $\vv = (\vvb^i, \vv^j)$ defined by 
\begin{align*} \vvb^i  &\= \frac{1}{\a_0} \left((-1)^{i +1} \hskip - 0.5cm \sum_{\smallmatrix 0 \leq J\leq |k - k_o -1|\\ 1 \leq j \leq n\endsmallmatrix} \a^J_{i|j} \left( \frac{d}{d x^1}\right)^{J_1}\!\!\!\!{\ldots} \left(\frac{d}{d x^m}\right)^{J_m}\!\!\!\left(\wh \vv^i \right) + (-1)^i h_i\right)\ ,\\
\vv^i &\=  y^i_{1_r} \vvb^r - \wh \vv^i\ .
\end{align*}
One can directly check that the $(m+n)$-tuple  $\vv = (\vvb^i, \vv^j)$ satisfies the equation 
\beq
(-1)^i \a_0 \vvb^i + \sum_{\smallmatrix 0 \leq J\leq |k - k_o -1|\\ 1 \leq j \leq n\endsmallmatrix} \a^J_{i|j} \left( \frac{d}{d x^1}\right)^{J_1}\!\!\!\!{\ldots} \left(\frac{d}{d x^m}\right)^{J_m}\!\!\!\left(\vv^i - y^i_{1_r} \vvb^r\right) = h_i\ . \eeq
From this, the  expressions \eqref{lillina} - \eqref{lillina2bis} together with  the assumptions $\l^{(\a)} =  \mu^{(\h)} = 0$ and  the identity \eqref{acciter},  one gets  that  $X_\vv$ satisfies 
\begin{align}
\label{ore12}Ê&\imath_{X_\vv} \a = \h|_\cU - \gz\ ,\\
\nonumber  \label{ore13} &   \imath_{X_{\vv}} \b\left(\frac{d}{d x^1},\ldots, \frac{d}{d x^m} \right) =   \left(\vv^i- y^i_{1_r} \vvb^r\right) \b_i =  -  \sum_{i = 1}^n \wh \vv^i \b_i  = \\
& \qquad \qquad \qquad \qquad \qquad = d  (- \h|_\cU+ \gz)\left(\frac{d}{d x^1},\ldots, \frac{d}{d x^m} \right)\ . 
 \end{align}
Hence \eqref{ohoh} and \eqref{ohoh1} hold. It remains to show that also equations \eqref{ohoh2} are satisfied. To show this, we observe that from  \eqref{ore12} and the construction of the $(m-1)$-form $\gz$, the $(m-1)$-form $\imath_X \a$ Êis an $m$-form of order less than or equal to $2k_o -k_\b$. Hence $d(\imath_X \a)$ is an $m$-form of order less than or equal to $2 k_o - k_\b + 1 \leq 2 k_o + 1$. Since  $k_o \leq \left[\frac{k}{2}\right]-1$, it follows that $d(\imath_X \a)$ is of order less than or equal to $k -1$ and  that  the contraction of  $d(\imath_X \a)$  with any vector field $\frac{\p}{\p y^j_J}$ with $|J| = k$ is $0$. From this,  \eqref{ohoh2} follows. 
\end{pf}
\bigskip
\section{An example of a form of Poincar\'e-Cartan type}
\label{Examples}
\setcounter{equation}{0}
 Let  $\cE=\bR^{1,3}$ be the space-time of Special Relativity and  denote  by $(x^0, x^1, x^2, x^3)$ and 
$$\h:= \h_{ij} dx^i \otimes dx^j\ ,\qquad \text{with}\ \ (\h_{ij}) = \left(\begin{array}{cccc} 1 & 0 & 0 & 0\\  0 & -1 & 0 & 0\\ 0 & 0 & -1 & 0\\ 0 & 0 & 0 & -1 \end{array}\right)$$ 
the standard coordinates and the standard flat metric of $\bR^{1,3}$, respectively
(as usual, we  follow the classical Einstein convention on summations). In Special Relativity  the electromagnetic field is represented by a closed  $2$-form  $\bF=F_{ij}dx^i\wedge dx^j$, that is a $2$-form which can be locally written as 
\beq\label{cem}\bF=d\bA=\left(\frac{\p A_j}{\p x^i}-\frac{\p A_i}{\p x^j}\right)dx^i\wedge dx^j\ ,\eeq
for a $1$-form $\bA=A_0dx^0+A_1dx^1+A_2dx^2+A_3dx^3$,  called  {\it $4$-potential} .  Since $\cE = \bR^{1,3}$ is contractible,  we may assume that $\bA$ is globally defined and  consider  the Maxwell  equations  in the vacuum as partial differential equations of second order on  the $4$-potential  $\bA$. It is well known that these equations  are precisely the Euler-Lagrange equations  for   the Lagrangian 
\beq \label{Lagrangiana4}L:J^1(T^*\cE)\lra \bR\ ,\quad L(j^1(\bA))\=-\frac{1}{16 \pi c}{|\bF|}^2_{\h}=-\frac{1}{16 \pi c}\h^{i\ell}\h^{jm}F_{ij}F_{\ell m} ,\eeq
where $F_{ij}$ are the coordinate components of  $\bF=d\bA$,  $(\h^{ij}) = (\h_{\ell m})^{-1}$ and $c$ is the physical constant  given by  the  speed of light.\par
\smallskip
Let $E \= T^* \cE$ and denote by   $\wh \xi^{(2)} = (x^i, A_m, A_{m,n}, A_{m, nr})$ a (global) system of 
adapted coordinates on the second order jet space $J^2(E) = J^2(T^* \cE)$.  If we denote by $\a_L$ the $4$-form on $J^2(E)$ 
$$\a_L = L(x^i, A_m, A_{m,n}) dx^0 \wedge dx^1 \wedge dx^2 \wedge dx^3$$
with $L$  defined in \eqref{Lagrangiana4}, we have that 
{\it a $1$-form $\bA: \cE \simeq \bR^4\longrightarrow T^* \cE$   satisfies  Maxwell's equations if and only if it satisfies the variational principle of the action $\cI_{[\a_L]}$}. \par
\medskip
The   proof of  Prop. A2 in \cite{Sp} (see also \cite{Ge}, Thm.1.3.11) gives an algorithm to   determine   $m$-forms of  Poincar\'e-Cartan  type in a given variational class. For the variational class  $[\a_L]$, such algorithm produces the $4$-form
\begin{multline}\label{alfetta}  \a=Ldx^0\wedge \ldots \wedge dx^3+\sum_{h=0}^3 {(-1)}^h\frac{\p L}{\p A_{k,h}}dx^0\wedge \ldots \underset{h}{\wh{}}\ldots \wedge dx^3\wedge \left(dA_k-A_{k,r}dx^r\right){=}\\
{=}\left(\!\frac{1}{16\pi c}F^{kh}F_{kh}\!\right)dx^0\wedge \ldots \wedge dx^3-\sum_{h=0}^3 {(-1)}^h\frac{1}{4\pi c}F^{kh}dx^0\wedge \ldots \underset{h}{\wh{}}\ldots \wedge dx^3\wedge dA_k.
\end{multline}
One can directly check that the differential  $d \a$ is a source form  (modulo a holonomic $5$-form) whose components are precisely the terms of the Euler-Lagrange equations of  $L$.\par
 \smallskip
 As is well known, Maxwell equations are conformally invariant. This corresponds to the fact that, for each $k \geq 1$, and  for  each  conformal Killing vector field  $X$ of $\bR^{1,3}$, the  corresponding vector field $\wh X^{(k)}$ on $J^k(T^* \cE)$, whose local  flows $\Phi^{\wh X^{(k)}}_t \in \text{Diff}_{\text{loc}}(J^k(T^* \cE))$ are the natural   lifts  of the local flows $\Phi^X_t \in \text{Diff}_{\text{loc}}(\cE)$ of $X$, is an infinitesimal symmetry for the action $\cI_{[\a]}$. For instance,  each   vector field $\x_{(j)} := \frac{\p}{\p x^j}$, $ 0 \leq j \leq 3$,  generating the  translations in the directions of the $x^j$-axis,  is clearly a conformal Killing vector field and  its associated vector field $\wh \x^{(1)}_{(j)} := \frac{\p}{\p x^j} + A_{j, k} \frac{\p}{\p A_k}$ on $J^1(T^* \cE)$ is an infinitesimal symmetry for   $\cI_{[\a]}$. One can directly check that (see e.g. \cite{Ge})
 \begin{multline} \imath_{\wh \x^{(1)}_{(j)}} \a = \imath_{V_{(j)}} d x^0 \wedge \ldots \wedge dx^3\ \quad \text{mod. holonomic  $3$-forms,} \\
 \text{with} \  V_{(j)} = \frac{1}{4 \pi c}\left(F_{j \ell} F_{km} \h^{\ell m} - \frac{1}{4} \h^{r\ell}\h^{sm}F_{rs}F_{\ell m}\right)\ .\end{multline}
Further, the components of $\imath_{V_{(j)}} d x^0 \wedge \ldots \wedge dx^3$ coincide (up to sign) with the components $T_{jr}$, $0 \leq r \leq 3$,  of the  electromagnetic stress-energy tensor $T$.  Hence,  from this and Theorem \ref{directNoether}, one has an alternative derivation of  the following well-known property:  the (equivalence classes of the) translational symmetries $\wh \x^{(1)}_{(j)}$, $0 \leq j \leq 3$, correspond via the Noether Theorem to the (equivalence classes of) the  conserved  stress-energy currents $\Phi_{(j)}:= T_{j  s}$.\par
We now recall that in \cite{AP}, Anco and Pohjanpelto classified all local conservation laws of Maxwell equations. There the authors proved that, modulo equivalences, 
any local  conservation laws is a linear combination of  some special currents,   constructed   using conformal Killing vector fields and conformal Killing-Yano tensor fields.  Using our proof of  Theorem \ref{inverseNoether}, one can determine the infinitesimal symmetries,  which correspond to all such conservation laws through a  contraction with the  form $\a$ of  Poincar\'e-Cartan type. From previous observations,  it is reasonable to expect that such infinitesimal symmetries (and,  consequently, most geometric properties of the $3$-form \eqref{alfetta}) are strongly related with the conformal Killing vector fields and, more interesting, with  the conformal Killing-Yano tensor fields 
of $\bR^{1,3}$. Making these relations explicit would very likely  pave the way towards  generalisations of various kind, quite useful for studying  for instance    Maxwell  equations  in curved spaces. \par
\bigskip
\appendix
\section{Erratum to ``Lie algebras of conservation laws of  variational  ordinary differential equations''}

The purpose of this short appendix  is  to remove an incorrect claim of   \cite{FS}. There,   in Prop. 3.5, it is improperly  stated that, when $\dim M \geq 2$, the $\cD$-symmetries coincide with the vector fields of the form $X = X_\vv$ with $\p \vv/\p y^i_{(k)} = 0$. The correct claim is that {\it the former are only a subset of the latter}. \par
\medskip
This does not effect the  results of the paper, provided that one  considers  {\it weak $\cD$-symmetries} (see Definition \ref{defDsymmetries} above) in place of  {\it $\cD$-symmetries}. We remark  that   weak $\cD$-symmetries can be considered  as  truncations up to order $k$  of    $\cD$-symmetries  of  $J^\infty(E)$ and that there is no difference between the two notions if one works on $J^\infty(E)$ in place of $J^k(E)$.  \par
\medskip
The correction  imposes  a few other minor  adjustments, which one can immediately  determine by  looking at  the more general results of  the present  paper. For instance,  the hypothesis  of Thm. 3.10 in \cite{FS} on the orders of  conservation laws and the $1$-form of Poincar\'e-Cartan type should be modified according to Theorem \ref{inverseNoether} of this paper, which  includes and extends the previous.   \par

\end{document}